\documentclass[journal]{IEEEtran}
\IEEEoverridecommandlockouts
\usepackage{epsfig,times}
\usepackage{subfigure}
\usepackage{multirow}
\usepackage{epsfig}
\usepackage{graphicx}
\usepackage{url}
\usepackage{times}
\usepackage{subfigure}
\usepackage{acronym}
\usepackage{psfrag}
\usepackage{booktabs}
\usepackage{amsmath}
\usepackage{array}
\usepackage{amssymb}
\usepackage{cite}
\usepackage{epsfig,times}
\usepackage{subfigure}
\usepackage{multirow}
\usepackage{epsfig}
\usepackage{graphicx}
\usepackage{url}
\usepackage{times}
\usepackage{subfigure}
\usepackage{acronym}
\usepackage{psfrag}
\usepackage{booktabs}
\usepackage{amsmath}
\usepackage{array}
\usepackage{amssymb}
\usepackage{fontenc}
\usepackage{booktabs}
\usepackage{color}
\usepackage{hhline}
\usepackage{soul}
\usepackage{tabu}
\usepackage{makecell} 
\usepackage{comment}

\usepackage{times,epsfig,subfigure,graphicx,acronym, footmisc}

\begin{document}
 \acrodef{ADC}{analog-to-digital converter}
 \acrodef{MADC}{multiplying analog-to-digital converter}
 \acrodef{SC}{switched-capacitor}
 \acrodef{DSP}{digital signal processor}
 \acrodef{S/H}{sample-and-hold}
 \acrodef{CDS}{correlated double sampling}
 \acrodef{PSNR}{peak signal-to-noise ratio}
 \acrodef{SNR}{signal-to-noise ratio}
 \acrodef{FPN}{fixed-pattern noise}
 \acrodef{CCI}{channel charge injection}
 \acrodef{DWT}{discrete wavelet transform}
 \acrodef{DCT}{discrete cosine transform}

 \date{}

\title{A 21.3\%-Efficiency Clipped-Sinusoid UWB Impulse Radio Transmitter with Simultaneous Inductive Powering and Data Receiving}

\author{\IEEEauthorblockN{Nima Soltani, \IEEEmembership{Member,~IEEE,} Hamed M. Jafari, \IEEEmembership{Member,~IEEE,} Karim Abdelhalim, \IEEEmembership{Member,~IEEE,} Hossein Kassiri, \IEEEmembership{Member,~IEEE,} Xilin Liu, \IEEEmembership{Senior Member,~IEEE,} and Roman Genov, \IEEEmembership{Senior Member,~IEEE}}\\
	

\thanks{N. Soltani and H. M. Jafari are with Intel Canada, North York, ON, M3C 3G8, Canada; K. Abdelhalim is with Sitrus Technology Corp, Irvine, CA, 91436, United States; H. Kassiri is with the Department of Electrical Engineering and Computer Science, York University, Toronto, ON, M3J 1P3; X. Liu (corresponding author xilinliu@ece.utoronto.ca) and R. Genov (roman@eecg.utoronto.ca) are with the Department of Electrical and Computer Engineering, University of Toronto, Toronto, ON, M5S 3G4, Canada.}
}

\markboth{IEEE Transactions on Biomedical Circuits and Systems, Accepted for Publication}{A 21.3\%-Efficiency Clipped-Sinusoid UWB Impulse Radio Transmitter with Simultaneous Inductive Powering and Data Receiving}

\maketitle

 \begin{abstract}
An ultra-wide-band impulse-radio (UWB-IR) transmitter (TX) for low-energy biomedical microsystems is presented. High power efficiency is achieved by modulating an LC tank that always resonates in the steady state during transmission. {\color{black}A new clipped-sinusoid scheme is proposed for on-off keying (OOK)-modulation, which is implemented by a voltage clipper circuit with on-chip biasing generation.} 
The TX is designed to provide a high data-rate wireless link within the 3-5 GHz band. The chip was fabricated in 130nm CMOS technology and fully characterized. State-of-the-art power efficiency of 21.3\% was achieved at a data-rate of 230Mbps and energy consumption of 21pJ/b. A bit-error-rate (BER) of less than 10$^{-6}$ was measured at a distance of 1m without pulse averaging. In addition, simultaneous wireless powering and VCO-based data transmission are supported. {\color{black}A {\color{black}potential} extension to a VCO-free all-wireless mode to further reduce the power consumption is also discussed.}

 \end{abstract}

\begin{IEEEkeywords}
Ultra-wideband (UWB) transmitter, impulse radio, high efficiency, low-power wireless, inductive powering, simultaneous power and data transfer
\end{IEEEkeywords}
\IEEEpeerreviewmaketitle

 \section{Introduction}
 \label{intro}
\begin{figure}[!t]
\centering
\includegraphics[width=1\columnwidth]{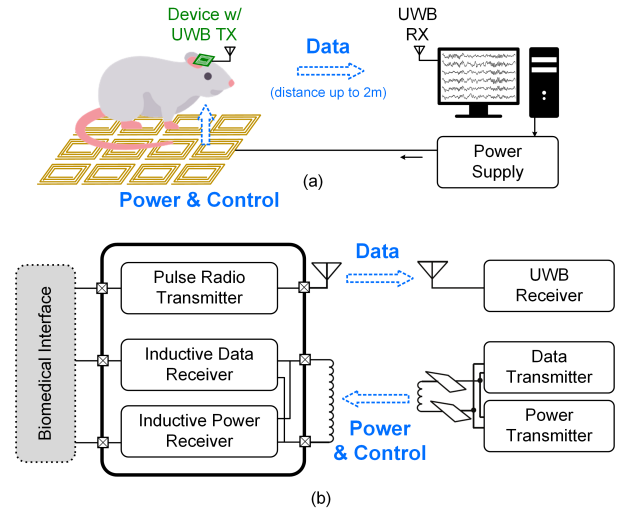}
\caption{(a) Illustration of a high-data-rate wireless TX link with simultaneous {\color{black}power receiving and data transmission for remote electrophysiological brain monitoring}. (b) Simplified block diagram of the proposed system (the biomedical interface is not included in this work).}
\label{fig_overview}
\end{figure}

Recent advances in biomedical devices have created demands for short-range high data-rate wireless transmission. As biomedical devices are integrating versatile sensors with higher spatial-temporal resolutions, the volume of data that needs to be streamed outside the device increases accordingly. Examples of such high data-rate biomedical devices include electronic neural interfaces \cite{song2022,zhang2020,liu2019wireless}, {\color{black}the targeted application of this work, as well as} retinal prosthetic implants \cite{Akinin2021}, wireless biomedical sensors \cite{liu2018}, and optical imaging devices \cite{Gagnon_Turcotte2018}. {\color{black}Fig. \ref{fig_overview} (a) illustrates an example of an electronic neural interface for chronic neuroscience studies in rodents, where a cellular inductive powering floor delivers energy to a freely-moving rodent for remote electrophysiological brain monitoring experiments \cite{Nima2016}. A simplified block diagram of such a wireless system is depicted in Fig. \ref{fig_overview} (b).}

A major challenge in designing high data-rate biomedical devices {\color{black}in general, and brain implants in particular,} is the limited power budget. The power source of such devices is either a lightweight low-capacity battery or a wireless power link by means of inductive couplingor energy harvesting \cite{liu2016}. The heat density requirement of miniaturized implantable devices {\color{black}adds} additional limitations. As a result, the available power of these devices is often limited to a few milliwatts per centimeter square, which is not sufficient for most conventional transmitters (TXs) to support the required data-rate \cite{Nima2016}.

Ultra-wideband impulse radio (UWB-IR) is one of the most suitable architectures for short-range ($<$10m) high data-rate ($>$10Mb/s) transmission \cite{singh2021,kim2016,Allebes2021,Lee2021,liu2017wireless}. A UWB-IR TX directly radiates a train of short pulses ($<$1ns). The direct transmission of short pulses results in a high data-rate as the symbol period can be as small as the duration of an individual pulse.

Compared with the state-of-the-art low-power narrow-band transmitters (e.g., Bluetooth or Zigbee \cite{narrow1,narrow2}), UWB-IR TX designs typically offer $\times$10 bandwidth at a per-bit energy dissipation that is orders of magnitude smaller \cite{crepaldi}. However, the output power efficiency of existing UWB-IR TX architectures is often low \cite{liu2021}. This is mainly due to the poor power efficiency of the output stage that drives the antenna \cite{kim2016,Allebes2021}. 
As a result, UWB-IR TXs are often designed with limited operating distance, which makes them unsuitable for many practical applications.

\begin{figure}[!t]
\centering
\includegraphics[width=1\columnwidth]{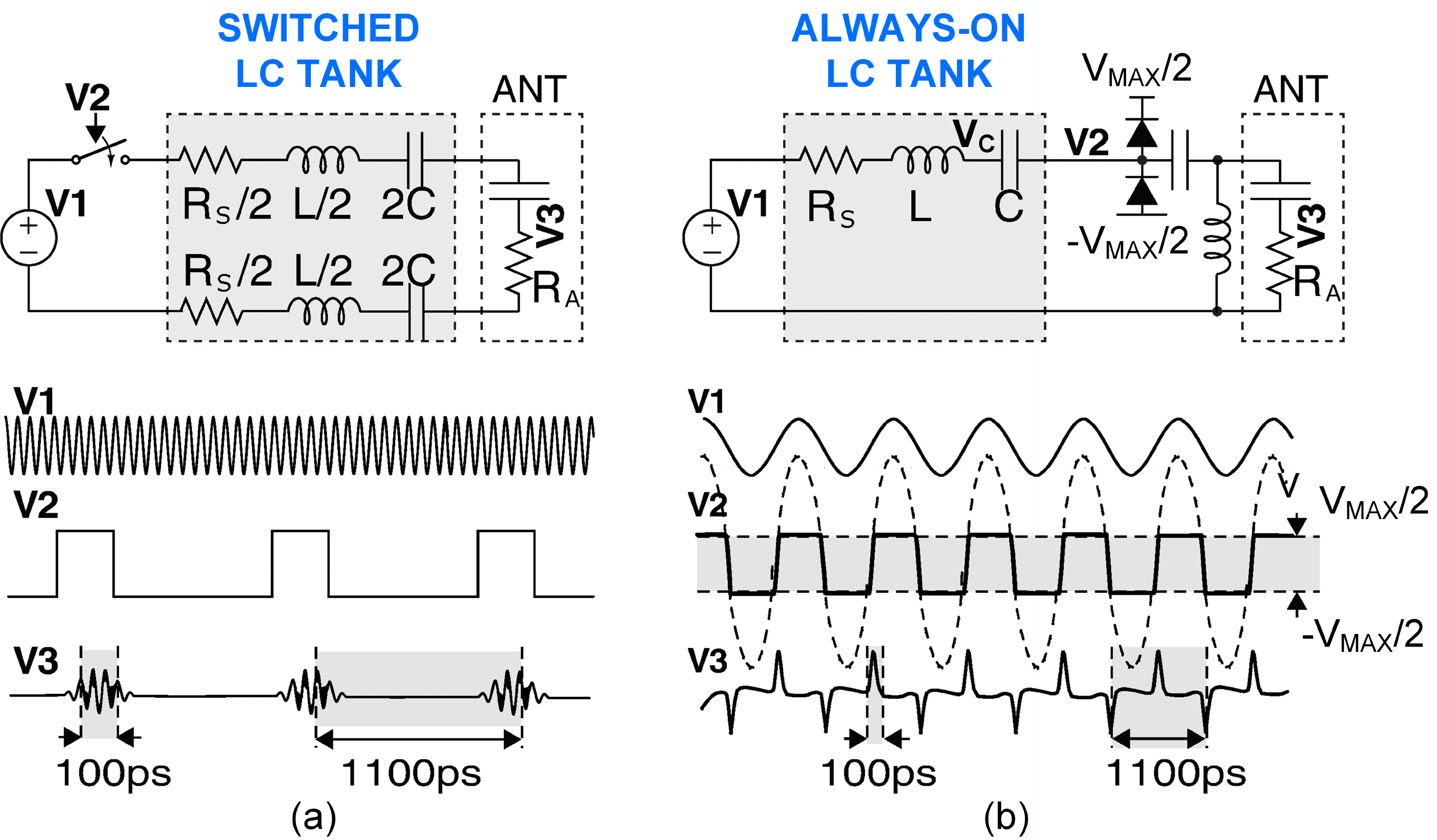}
\caption{{\color{black}(a) }Simplified diagrams of the conventional and {\color{black}(b)} the proposed pulse generation using a high-frequency transient and a low-frequency steady state LC tank, respectively. {\color{black}The proposed design avoids the energy-costly on-off tank transitions.}}
\label{figscheme}
\end{figure}

Several innovative low-power UWB-IR architectures have been proposed \cite{kassiri2017,Majid,chen,geng,streel2017,lee2019,kopta2019,Rahmani2021,Mirbozorgi2016,liu2017fully}. The designs in \cite{kassiri2017, Majid} use a combination of CMOS inverters {\color{black}as delay lines in order to} generate the UWB waveform and drive the antenna. {\color{black}Theoretically,} the TX only radiates power during the logic-state transitions {\color{black}and thus} the power is limited by the rising and falling times of the inverters. {\color{black}In practice, due to} the additional power consumption overhead from the digital delay lines and pulse-shaping circuits, the resulting overall TX power efficiency is often poor \cite{singh2021}.

On the other hand, the designs in \cite{chen,geng} generate UWB pulses by turning on and off a digitally-controlled cross-coupled LC oscillator, as illustrated in Fig.~\ref{figscheme} (a). In these cases, the TX efficiency is limited by the startup time of the oscillator. Because of the lower voltage swing, the oscillator's power efficiency during startup is much lower than its steady state. In addition, the oscillation frequency in these designs often needs to be several times larger than the pulse bandwidth, which makes the oscillator further more power-hungry.

If the UWB pulses can be generated from an LC oscillator operated in the steady state, the LC tank can always resonate with high power efficiency \cite{nima2015}. Moreover, by eliminating the need for periodically turning on and off the LC tank, the bitrate is no longer limited by the startup time of the LC tank, thus a more power-efficient LC tank with a lower resonant frequency can be employed.

{\color{black}This paper introduces a {\color{black}new} UWB-IR TX architecture that achieves a state-of-the-art TX power efficiency. The proposed design has three key {\color{black}advantageous} features:
\begin{itemize}
    \item First, a novel clipped-{\color{black}sinusoid} pulse generation scheme is proposed, which integrates a voltage clipper circuit at the output of an LC tank that is always in the steady-state, {\color{black}thus avoiding the energy-costly on-off transitions}.
    \item Secondly, the UWB-IR transmitter can be inductively powered (and can receive control commands) while {\color{black}simultaneously} transmitting data.
    \item Thirdly, by coupling the TX inductor with the power receiving coil, the frequency synthesizer circuit can be eliminated to {\color{black}enable a VCO-free all-wireless mode. This mode can further reduce the power consumption and is especially suitable for {\color{black}ultra-}low-power biomedical sensors.}
\end{itemize}
}

{\color{black}A part of this design and preliminary experimental results have been {\color{black}briefly} reported in \cite{nima2015}. In this paper, we {\color{black}expand on that report and} present the following {\color{black}additional} aspects of the work that {\color{black}have not been} {\color{black}previously} covered:
\begin{itemize}
\item A detailed analysis of the design methodology for improving the TX efficiency and {\color{black}of its} circuit implementation and operation. 
\item Introduction {\color{black}and validation} of the simultaneous inductive powering and data {\color{black}transmission} scheme, which is an essential {\color{black}newly reported} feature of the system.
\item Discussion of the {\color{black}newly reported} all-wireless {\color{black}VCO-free} configuration {\color{black}with both power and clock received wirelessly}, which can {\color{black}further} simplify the circuit {\color{black}implementation} and reduce the power consumption. 
\item A detailed description of the experimental setup and {\color{black}additional} measurement results and {\color{black}a comparison} with state-of-the-art designs.
\end{itemize}
}

The rest of the paper is organized as follows. Section II analyzes the power efficiency in UWB-IR TX designs. {\color{black}Section III presents the new UWB-IR TX system and circuit implementation, as well as an all-wireless configuration supporting simultaneous powering and data transmission.} Section IV presents the detailed measurement results of the prototype system {\color{black}in various test modes}. {\color{black}Section V discusses an extension of this work to support a VCO-free all-wireless mode to further reduce the power consumption. Section VI compares the performance of the presented work with state-of-the-art UWB TX designs.} Finally, section VII concludes the paper.

\section{Improving TX Power Efficiency}
\label{TXefficiency}
Fig.~\ref{figscheme} (a) shows the operational principles of a conventional UWB-IR TX based on switching on and off a cross-coupled LC tank VCO. A UWB pulse is generated when the tank is turned on by the baseband signal $V_2$. Considering only the power loss in the LC tank and neglecting the loss in the active components of the VCO, the power efficiency can be approximated by:

\begin{subequations}
\begin{align}
\eta_{TX}&= \frac{R_A I(t)^2}{V_1I(t)} \nonumber\\
&=\frac{R_A\left[u(t_p-t)(e^{\alpha t}-1)-u(t-t_p)e^{-\alpha(t-t_p)}\right]}{R_A+Rs}\times 100\%,\nonumber
\end{align}
\label{eq:B21}
\end{subequations}

\noindent where $\alpha$ is the time constant equal to ${-L}/[2(R_S+R_A)]$, and $t_p$ is the pulse width (during which $V_2=1$ in Fig.~\ref{figscheme}(a)). Due to the highly underdamped response of the LC tank in this case ($\alpha \ll \omega_o$), the numerator is a small fraction of $R_A$, therefore $\eta_{TX}$ is limited by the transient behavior of the LC tank.

It should be noted that the above expression of power efficiency of the LC tank VCO-based UWB-IR TX is based on the assumption that the duration of signal $V_2$ is much smaller than the time constant of the tank, i.e. $e^\alpha t_p \ll 1$. This condition is critical because as the oscillation builds up and the swing of the signal $V_3$ increases, the cross-coupled transistors in the LC tank VCO become increasingly non-linear, while these non-linearities are not reflected in the expression above. 

In the actual implementation of the UWB-IR designs as described in \cite{chen,geng}, the condition $e^\alpha t_p \ll 1$ is valid since the pulse duration $t_p$ lasts only for a few oscillation cycles. This is while the number of oscillation cycles that would make the condition valid must be comparable to the ratio of the tank's resonant frequency $\omega_o$ to its neper frequency $\alpha$. Since this ratio is essentially equal to $2Q$, which is an order of magnitude larger than the number of oscillation cycles, the above approximation is considered valid.

Fig.~\ref{figscheme} (b) illustrates the operational principle of the proposed UWB-IR TX. Two pulses are generated in every oscillation period of the LC tank. These high-bandwidth pulses are generated by voltage clipping at the output of the LC tank {\color{black}by the} two diodes {\color{black}connected in series} between $V_{MAX}/2$ and $-V_{MAX}/2$. The clipped signal, $V_2$, contains higher-order harmonics due to the abrupt limiting action of the diodes. The spectral power of the higher-order harmonics depends on the threshold voltages {\color{black} $V_{MAX}$/2 and -$V_{MAX}$/2}, which {\color{black}can be} digitally {\color{black}set} by digital-to-analog converters (DACs). As shown in Fig.~\ref{figscheme} (b), the raw UWB pulse train, $V_3$, is created at the antenna by high-passing the clipped signal $V_2$. {\color{black}It should be noted that Fig. \ref{figscheme} (b) is a conceptual representation of the key idea, which excludes the OOK coding scheme. Each pulse in $V_3$ may actually represent multiple pulses in a ripple, depending on the quality factor of the LC tank and the clipping voltages.}

In the proposed UWB-IR TX architecture, since all the pulse power is sourced from the LC tank that is resonating in the steady state, the overall efficiency of the system is given by:
\begin{equation}
\eta_{TX}=\frac{R_A}{R_A+Rs}\times 100\%.\nonumber
\label{eq:eff}
\end{equation}
Therefore, the power efficiency of the TX remains high at all times, as long as the LC tank is implemented with a high-Q inductor, and the diode junctions are abrupt enough to extend the pulse bandwidth over the frequency band of interest.

\section{UWB-IR TX Design}

\subsection{TX System Architecture}
\label{efftrans}
\begin{figure}[!ht]
\vspace{1ex}
\centering
\includegraphics[width=1\columnwidth]{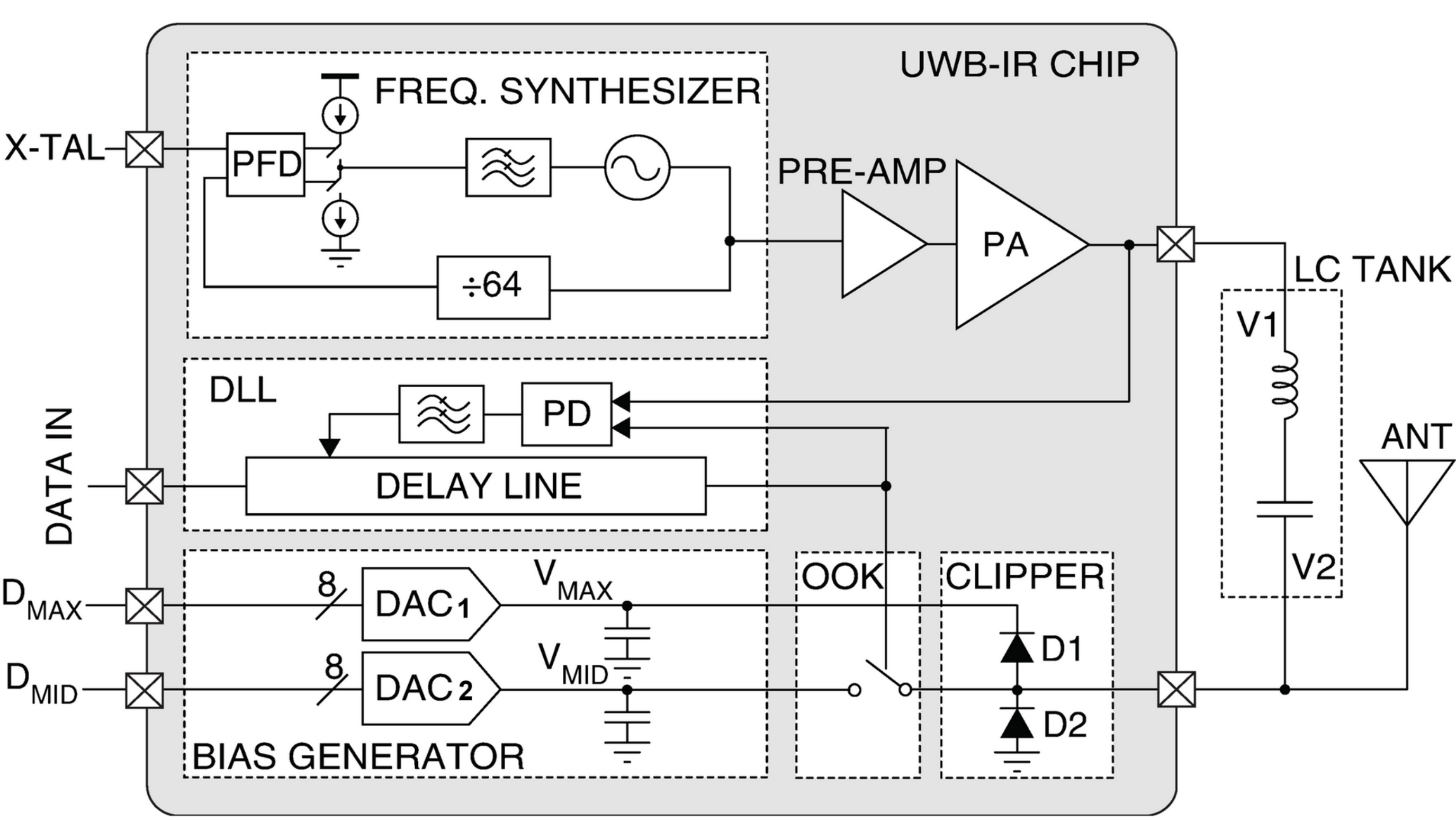}
\caption{Block diagram of the presented UWB-IR TX system.}
\label{figblock}
\end{figure}

Fig.~\ref{figblock} shows a detailed block diagram of the proposed UWB-IR TX system. It consists of a 915MHz frequency synthesizer, a pre-amplifier, a power amplifier (PA), a resonant LC tank, a two-diode clipper circuit, and a delay-locked loop (DLL) for the OOK pulse modulation. A 915MHz pure tone is generated by the on-chip frequency synthesizer from an off-chip 14.3MHz crystal. 
The frequency synthesizer uses a cross-coupled LC tank voltage-controlled oscillator (VCO) and a true single-phase clock (TSPC) flip-flop phase-frequency detector (PFD) \cite{kassiri2017}. 
The synthesizer is connected to a pre-amplifier which drives the inductive-load PA, which in turn drives a high-Q series LC tank. The LC tank is implemented off-chip.
The threshold voltage $V_{MAX}$ and the AC ground $V_{MID}$ {\color{black}(ideally equal to $V_{MAX}$/2)} are set by 8-bit DAC1 and DAC2, respectively. 
These voltages can be digitally adjusted for optimizing the radiated power. {\color{black}The DACs can also be used for compensating PVT variations.}

The UWB pulse train is modulated by switching the output of the LC tank based on the data stream. When shorting $V_2$ to the AC ground $V_{MID}$, both diodes remain off and no pulse is generated. Since the swing of $V_2$ remains many times larger than the swing of $V_1$ regardless of the switch state, the switching action does not impact the quality factor of the LC tank. The DLL is situated between the data stream and the switching node to ensure proper timing between the switching signal and the transitions of the diodes. 

\subsection{TX Circuit Implementation}
\label{implementation}
\begin{figure}[!ht]
 \centering
 \includegraphics[width=1\columnwidth]{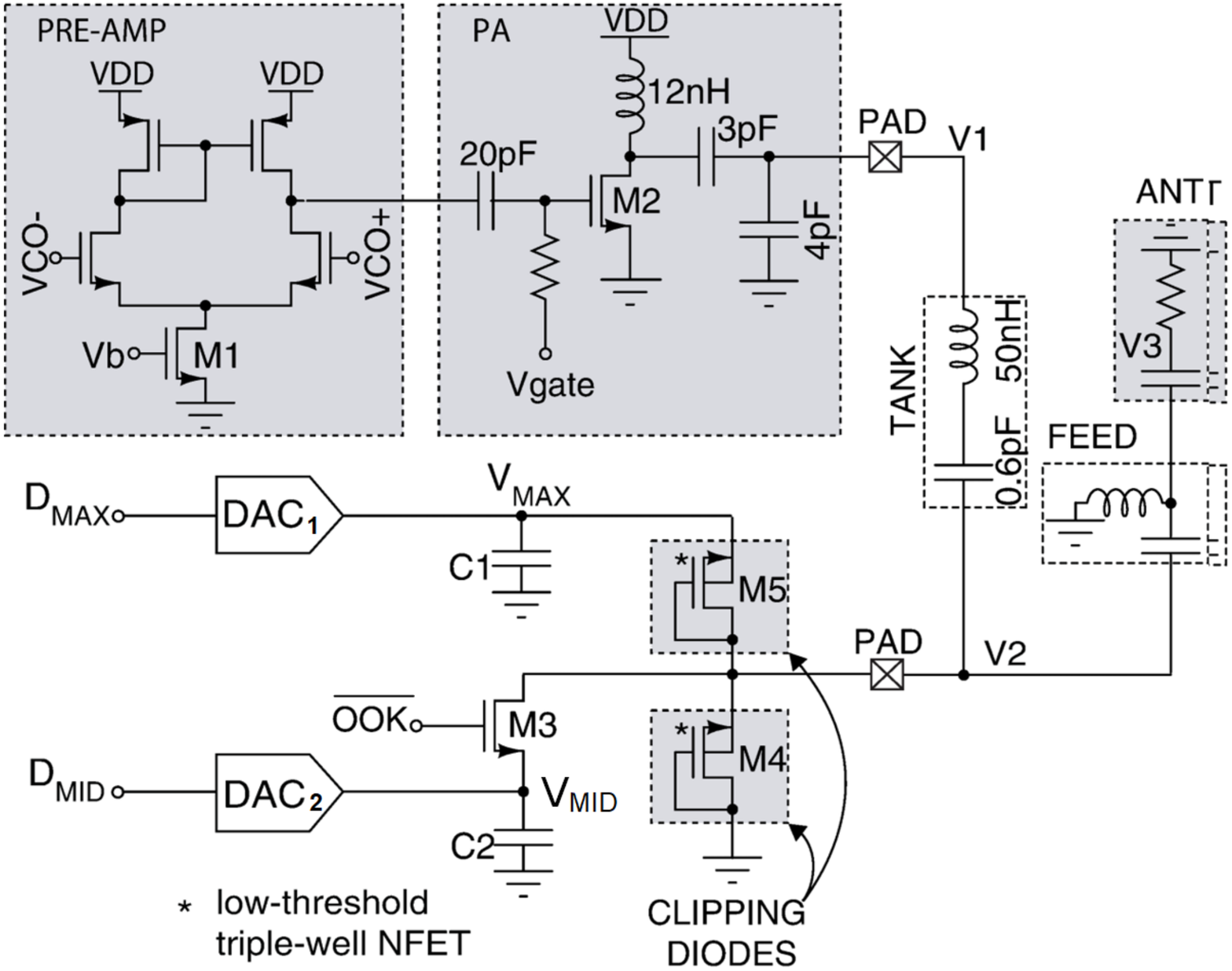}
 \caption{Simplified transistor-level schematic of the proposed UWB pulse generator.}
 \label{label_fig4}
\end{figure}

The proposed design is implemented in a standard 130nm CMOS technology. 
Fig.~\ref{label_fig4} shows the simplified schematic of the pulse generation circuit including the pre-amplifier, PA, LC tank, diode clippers, and the OOK switch. {\color{black}The pre-amplifier is a differential pair with a diode-connected current-mirror load. It is used to convert the differential VCO outputs to a single-ended output, and provide an additional gain for driving the PA. The choice of the PA architecture is important and requires careful consideration. While most of the advantage in the proposed design comes from the fact that the LC tank is always on and does not require energy-consuming on-off transitions, as was described in Fig. 2, the PA is another aspect of the design that can have a significant effect on energy efficiency. 
{\color{black}In this work, we adopted a Class-C PA, as depicted in Fig. \ref{label_fig4} (top, middle). Although Class-C PAs are not as power efficient as their Class-D counterparts, the amplifier efficiency in this work is maintained high as the PA is on only for a small fraction of the input signal period.} Class-C amplifiers are inherently nonlinear. Distortion is reduced by the tuned LC components at its output, {\color{black}eliminating the need for active filters at the cost of area}. The filter is then coupled to an off-chip LC tank connected to a chip antenna. Impedance variation in the external components such as in the bondwires and antenna have only a minor effect on the PA efficiency, as bondwire geometry is design-controlled and the antenna is packaged.
} 



The clipping diodes are implemented by two diode-connected triple-well NMOS devices $M4$ and $M5$, which have a width of $50\mu m$ and the minimum length. The OOK switch is implemented by an NMOS device $M3$, which has a width of $200\mu m$ and the minimum length. 
{\color{black} When transmitting "1", the $\overline{OOK}$ signal is low and the clipped signal $V_2$ is sent to the antenna; when transmitting "0", the $\overline{OOK}$ signal is high and $V_2$ is shorted to $V_{MID}$. When shorting $V_2$ to $V_{MID}$, both diodes remain off, therefore no pulse is generated. } $C_1$ and $C_2$ are used as decoupling capacitors at the outputs of DAC1 and DAC2, respectively. Each of them has a total capacitance of $200pF$. {\color{black}DAC1 and DAC2 can be individually programmed. Despite the fact that $V_{MID}$ is set to be half $V_{MAX}$ by default, using two DACs allows for more flexibility in choosing the parameters and potentially compensating for mismatches (e.g., the threshold variation of the two diodes). The matching of the two DACs is not a major concern since the output voltages can be calibrated and the 8-bit resolution is sufficient for the purpose of this design.}

\begin{figure}[!ht]
 \centering
 \includegraphics[width=1\columnwidth]{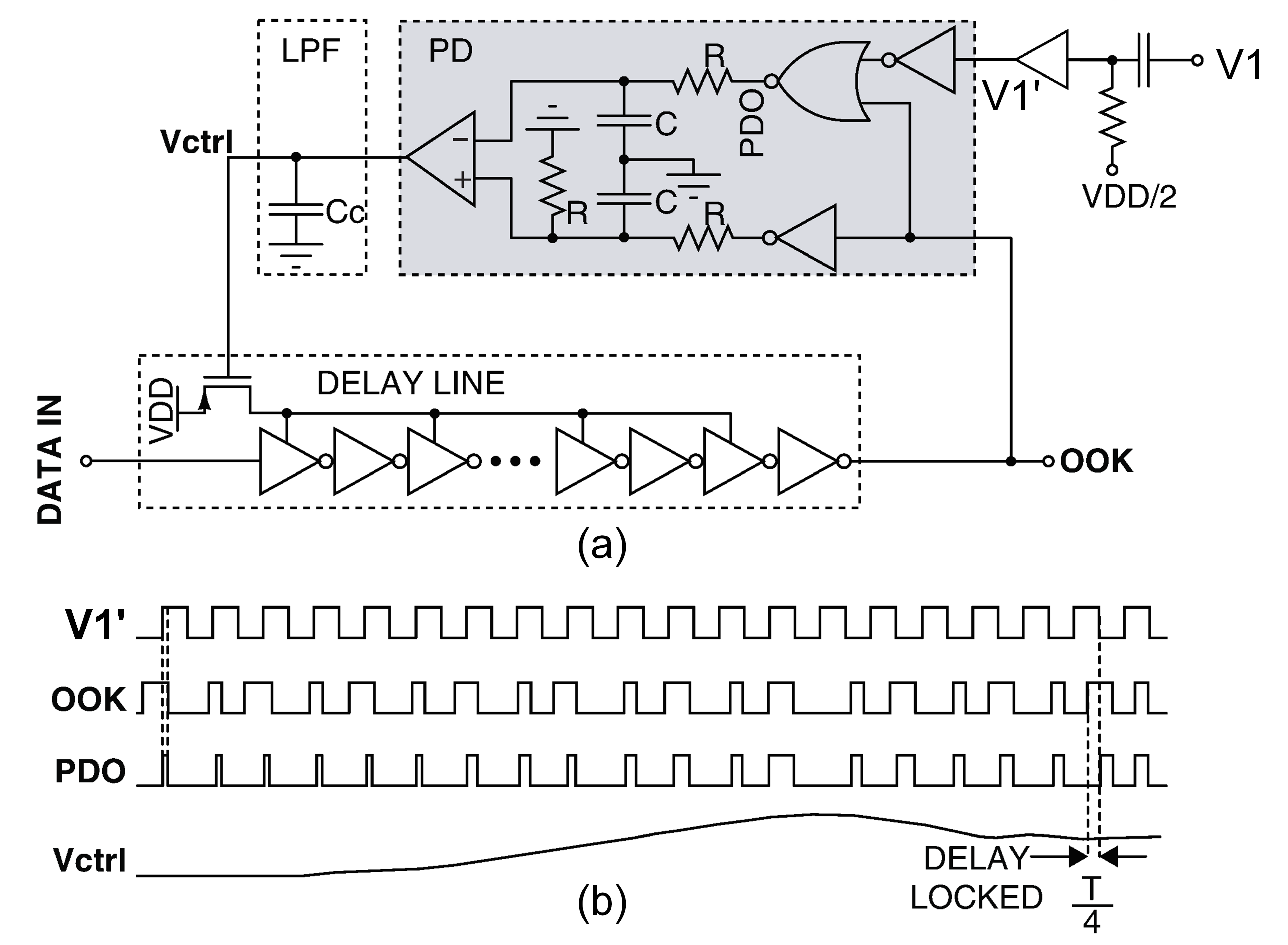}
 \caption{(a) Simplified schematic of the DLL circuit developed for the OOK modulation, and (b) {\color{black}a time-domain} illustration of its operation.}
\label{figOOK}
\end{figure}

Fig.~\ref{figOOK} (a) shows the schematic of the DLL circuit. 
The DLL is comprised of a phase detector (PD), a low-pass filter (LPF), and a variable delay line. The loop regulates the delay of the inverter-chain-based delay line until its output precedes the PA's output {\color{black}($V_1$)} by exactly $T/4$, where $T$ is the oscillation period of the LC tank. 
The DLL quantifies the misalignment between the two rising edges by comparing the duration of every OOK ``1" bit with the duration of the concurring ``1" bit at {\color{black}$V_1$}. To implement this, {\color{black}$V_1$ is first AC-coupled and digitally buffered to generate a rectangle-pulse signal $V_1'$. The PA's output is designed to not exceed the supply voltage at all output power levels, and a diode protection circuit was added to its output pad to avoid damaging the input gate of the digital buffer. $V_1'$ is then inverted and} NORed with the output of the delay line. Two pulse-averaging RC filters quantify the pulse widths of the data ``1" bits as seen at the output of the delay line and the NORed output. The RC filter for the delay line output has a DC gain of 1/2, such that these two filters' outputs are at equal levels when the PA and the data bits are exactly $T/4$ apart in phase. The difference between the outputs of the two RC filters is quantified by a differential amplifier which is implemented as a self-biased differential pair. A compensation capacitor $C_c$ is added to the differential amplifier's output, $V_{ctrl}$, to stabilize the feedback loop.

It should be noted that an auxiliary quarter-period ``1" bit is inserted after every data bit, as shown in Fig. \ref{figOOK} (b). This is to ensure that the loop settles only when the bits \textit{precede} the PA zero crossings by a $T/4$. Without the auxiliary ``1" bit, the DLL may settle falsely when the bits \textit{follow} the PA's output by a $T/4$.

\subsection{\color{black}VCO-based All-Wireless Mode}
\label{simtrans}

\begin{figure}[!ht]
\centering
\includegraphics[width=1\columnwidth]{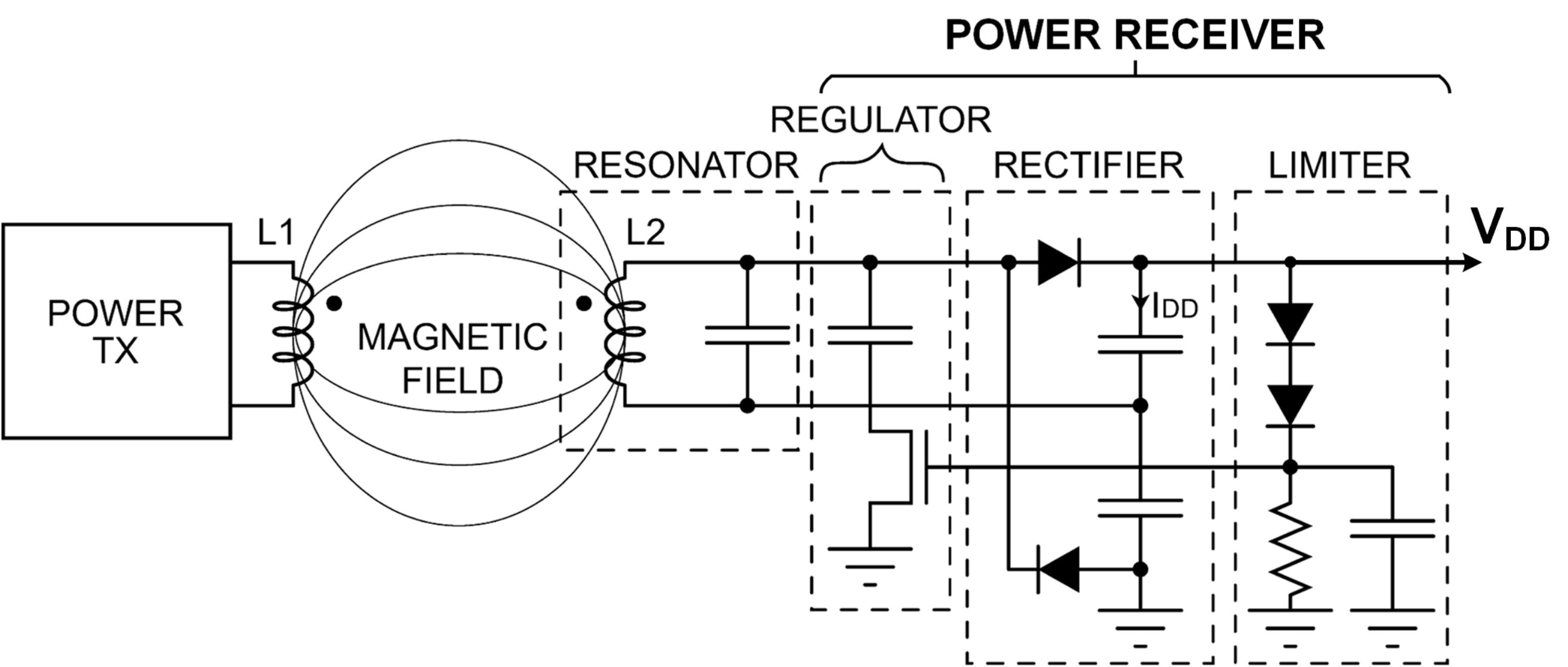}
\caption{\color{black}Simplified diagrams of the inductive power transfer sub-system for the all-wireless mode.}
\label{figsimul}
\end{figure}

{\color{black}The proposed TX system features a VCO-based all-wireless mode that supports simultaneous powering and data transmission. }
Fig.~\ref{figsimul} shows the block diagram of the inductive power receiver designed in this work to power the UWB-IR TX. 
The level of the received power is regulated by adjusting the tuning of the LC tank, such that the tank becomes detuned slowly when there is excess in the received power. {\color{black}The detuning may result in degradation in the overall power transfer efficiency, but the highest priority for power receiver design in applications such as implantable medical devices is to ensure sufficient power receiving for normal operation, while making sure the heat dissipation will not damage the tissue environment. The detuning helps reduce the excess heat from the LC tank. The efficiency of the power transfer link is a secondary concern in this case, since the power is generated by external devices with less energy constraints.} The RF power of the LC tank is transformed to DC using a dual-halfwave rectifier, which offers better power efficiency than a full-wave bridge rectifier. A limiter is used at the output of the rectifier to avoid a sudden surge in the rectified voltage. Slow-varying feedback is fed from the limiter to the regulator for keeping the output voltage level steady over a long time without dissipating excess power and generating unnecessary heat.

\section{Experimental Results}
\label{Result}

\subsection{{\color{black}IC, TX PCB, \& RX PCB Prototypes}}
\begin{figure}[!ht]
\centering
\includegraphics[width=.8\columnwidth]{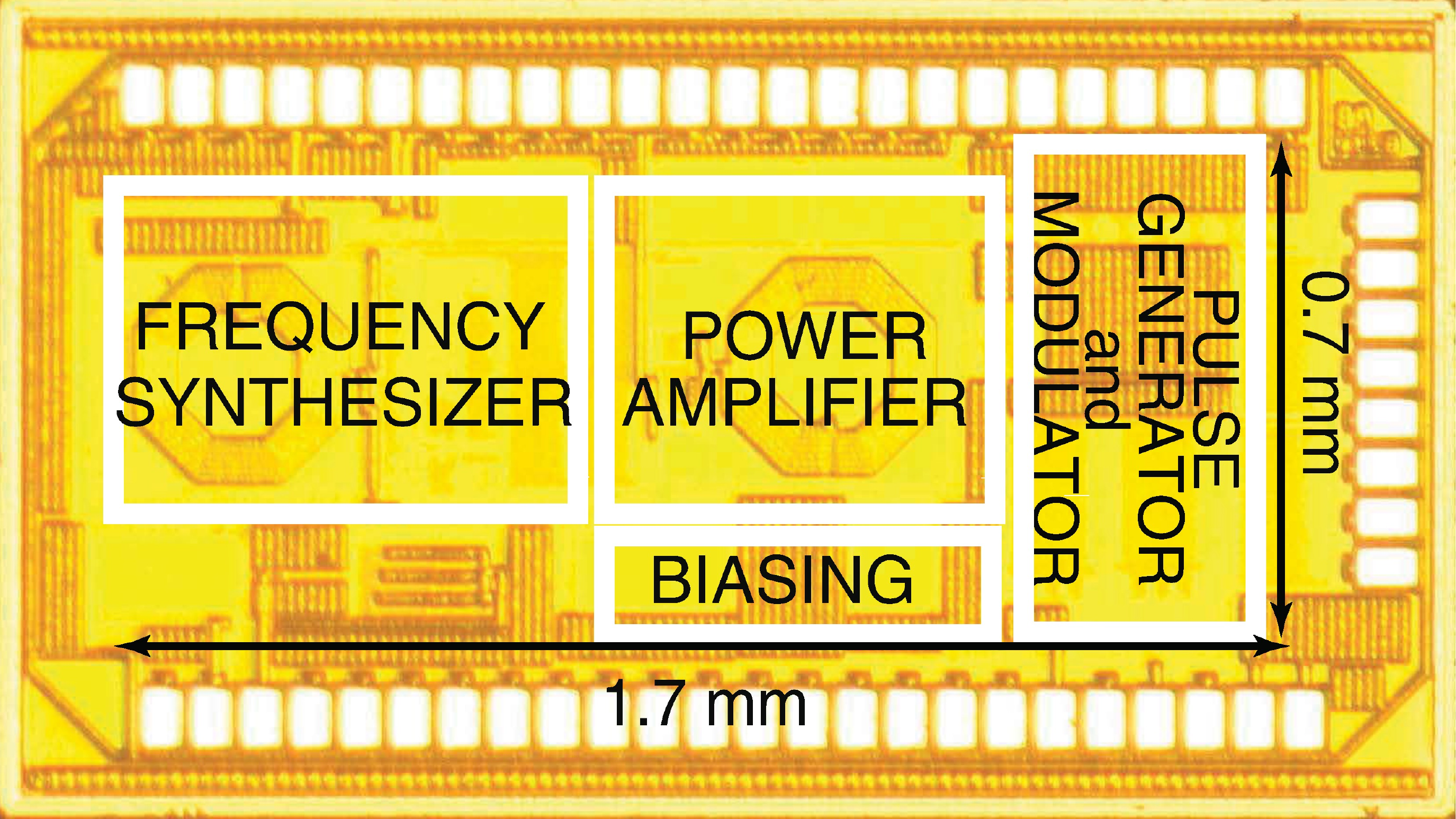}
\caption{Micrograph of the prototype chip fabricated in 130nm CMOS.}
\label{figcomp_die}
\end{figure}

The design was fabricated in a 130nm standard CMOS technology, occupying a silicon area of 1.7mm$\times$0.7mm. The micrograph of the fabricated chip is shown in Fig.~\ref{figcomp_die}. 

\begin{figure}[!ht]
  \centering
  \includegraphics[width=1\columnwidth]{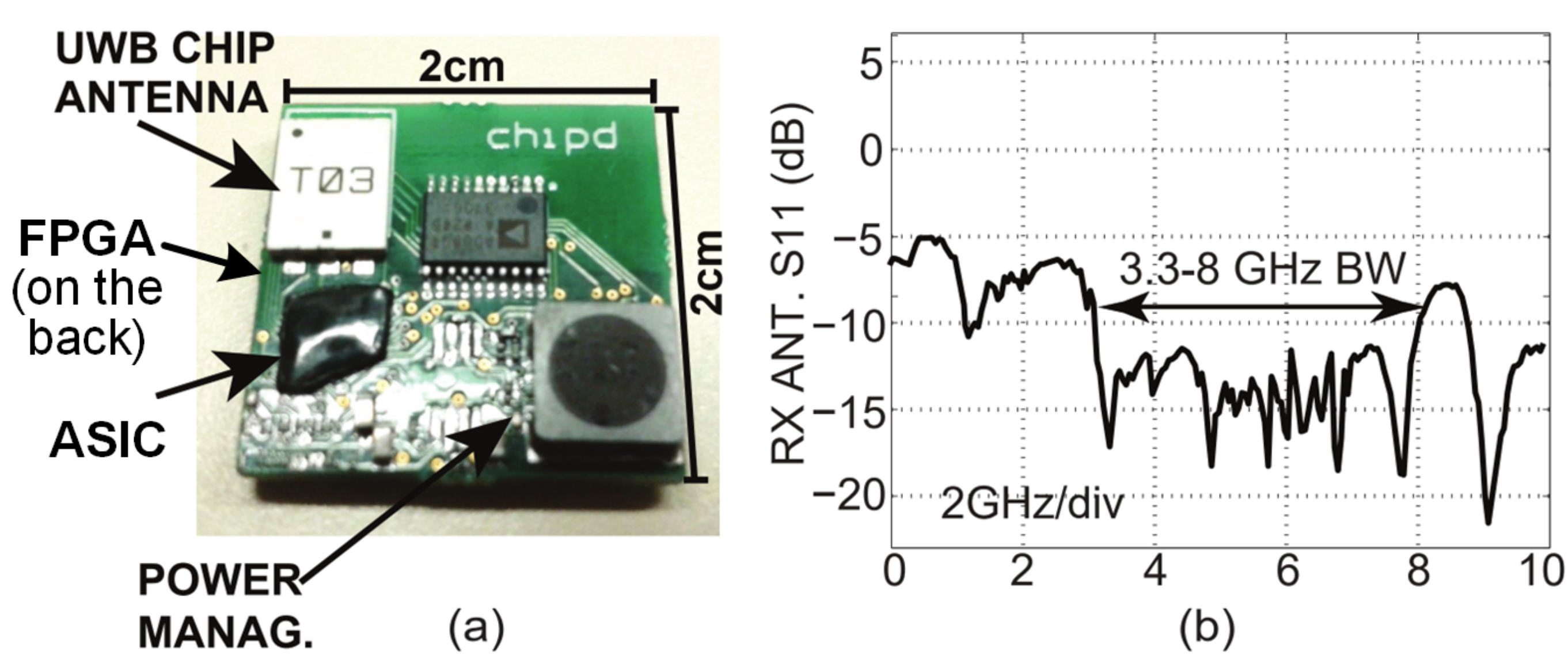}
  \caption{(a) The assembled UWB-IR TX board, and (b) the measured TX antenna return loss (S$_{11}$).}
  \label{figtx}
  \end{figure}

To validate the wireless operation of the designed UWB-IR chip {\color{black}in experimental neuroscience applications}, a mini-board was developed as shown in Fig.~\ref{figtx} (a). {\color{black}The mini-board has a dimension of 2cm by 2cm, which integrates the fabricated chip, a chip antenna, a low-power FPGA, and power management units. The size of the board permits many wearable and implantable applications \cite{zhang2020}, while maintaining flexibility and generality as needed to accommodate multiple applications.} 
The generated UWB pulses were radiated from the chip antenna assembled in the upper-left corner of the board. The chip antenna return loss is plotted in Fig.~\ref{figtx} (b), which verifies that the antenna radiates best within the 3.3GHz-8GHz band. In addition, an on-board power management block was integrated to rectify, down-convert, and regulate the high-power signal from the inductive coil connected to the board. {\color{black}For extreme volume-constrained applications, the device form factor can be further reduced by integrating the power management circuits and the FPGA digital logic circuits on the chip. This would support an even wider range of applications, such as implantation in the deeper brain \cite{Krauss2021}, but at the expense of reduced general-purpose utility.}

\begin{figure}[!ht]
 \centering
 \includegraphics[width=1\columnwidth]{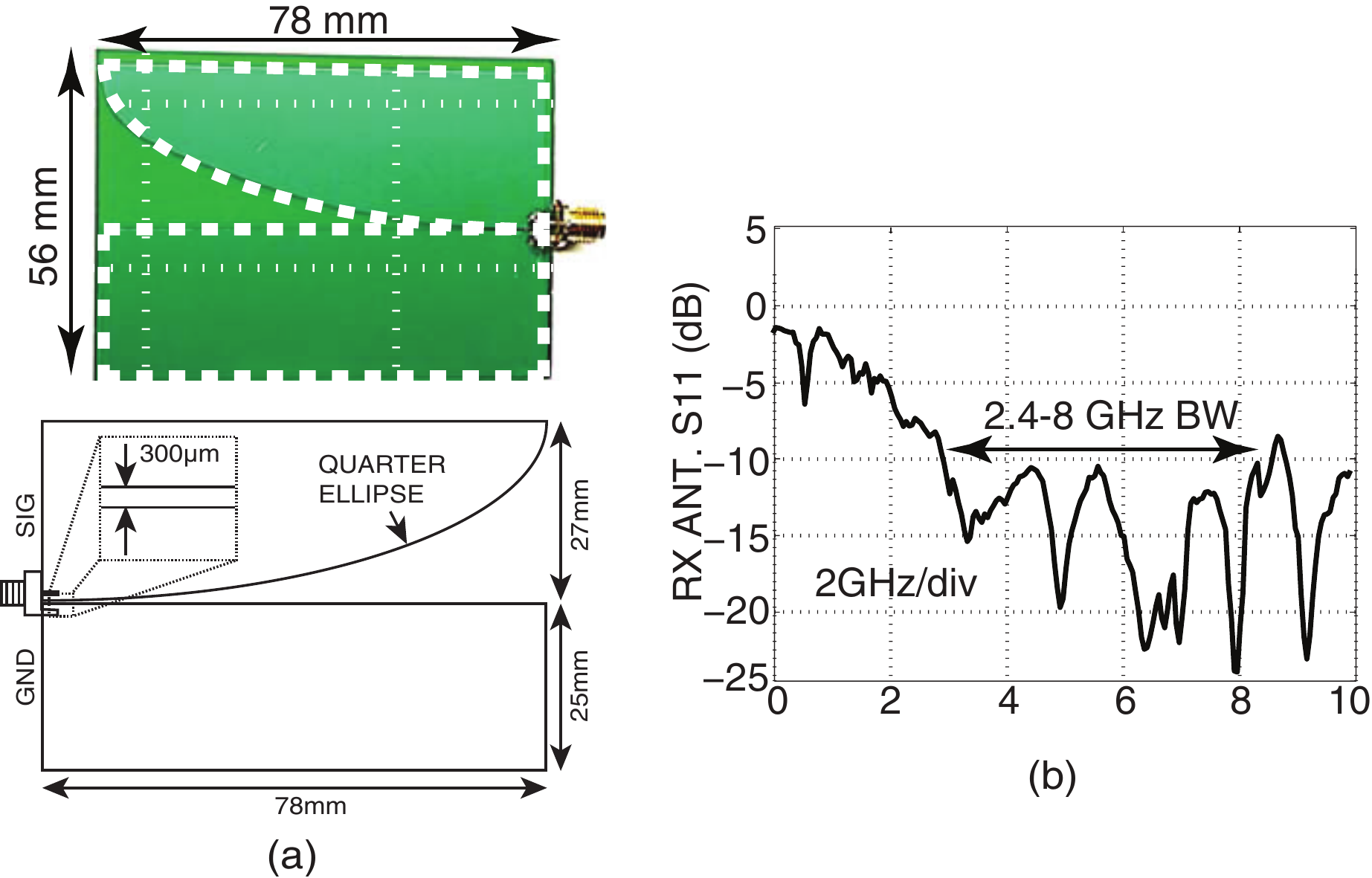}
 \caption{(a) The UWB-IR RX PCB antenna, and (b) the measured RX antenna return loss (S$_{11}$).}
 \label{figrx}
 \end{figure}

The transmitted UWB-IR pulses are picked up by an RX antenna, which is shown in Fig.~\ref{figrx} (a). {\color{black}The RX antenna was used only with the external receiver and was not designed to be integrated into the miniature module. Since there is no form factor constraint for the external RX antenna, it was designed to radiate best within 2.4-8 GHz and its size was set by the lower limit of the radiation frequency (i.e., 2.4GHz).}

\subsection{Wired-Power Wired-Data Test Mode}
\begin{figure}[!ht]
 \centering
 \includegraphics[width=1\columnwidth]{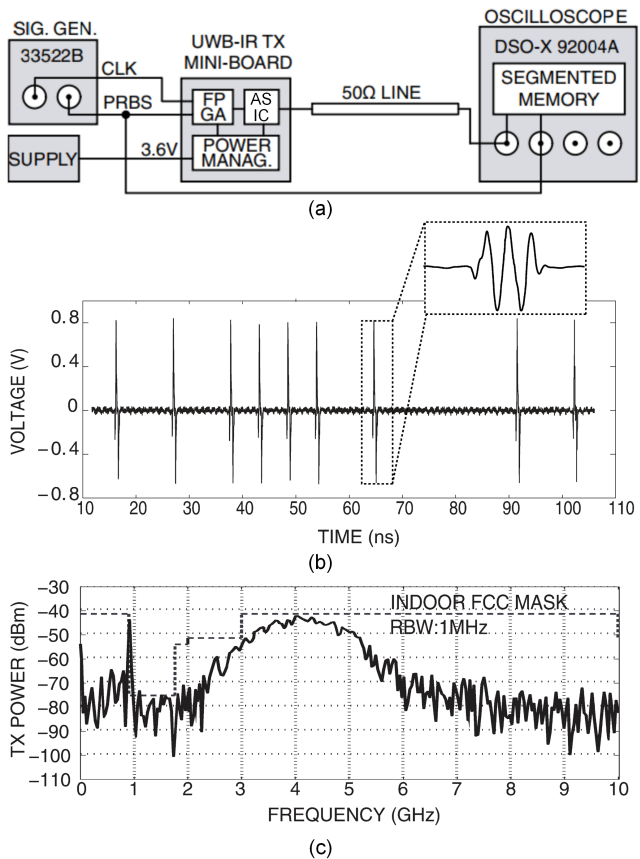}
 \caption{{\color{black}(a) Experimental setup for testing the developed UWB-IR TX output in the {\color{black}wired-power wired-data test} mode. (b) Measured modulated time-domain output of the TX, and (c) the corresponding power spectrum.}}
 \label{fig_exp1}
\end{figure}

{\color{black}We first characterized our design in a wired-power wired-data test mode. This test mode allows us to fully evaluate the functionalities and performance of the modulation and demodulation scheme and characterize the TX output power for ensuring its compliance with the standards approved by the Federal Communications Commission (FCC). In this test mode,} an external wired power supply was used to power the on-board regulators. The output of the TX was first measured directly using an Agilent DSO-X 92004A oscilloscope at a sampling rate of 80 GSa/s. The experimental setup is illustrated in Fig. \ref{fig_exp1} (a).
A {\color{black}low-power Actel} FPGA was used for chip configuration and data handling. The baseband data used for testing the wireless link was fed by a pseudorandom binary sequence (PRBS) generator. The PRBS sequence was simultaneously fed to another channel of the oscilloscope for post-processing and bit error rate (BER) computation.
{\color{black}The modulated output of the chip was measured at the maximum output power allowed by the FCC mask. }
{\color{black}Fig.~\ref{fig_exp1} (b)} shows the transient output of the TX modulated by a PRBS. {\color{black}Fig.~\ref{fig_exp1} (c)} shows the spectrum of the OOK modulated pulse train, which spreads over the 3GHz-5GHz frequency range. {\color{black}A spur at 915MHz is visible in the measured spectrum. We hypothesize that the spur is due to the coupling from PLL from the layout. A more careful layout review should be conducted in the future to avoid potential coupling effect. Nevertheless, the measured output spectrum of the design was under the UWB spectral mask approved by FCC and the spur didn’t affect the transmission performance.} At higher UWB frequencies, antennas either have a small aperture or are extremely sensitive to misalignment. Therefore extending radiated spectral power beyond this frequency range is of less interest, especially for wearable and implantable biomedical devices where misalignment of antennas cannot be totally avoided.

\subsection{\color{black}Wired-Power Wireless-Data Test Mode}
\begin{figure}[!ht]
 \centering
 \includegraphics[width=1\columnwidth]{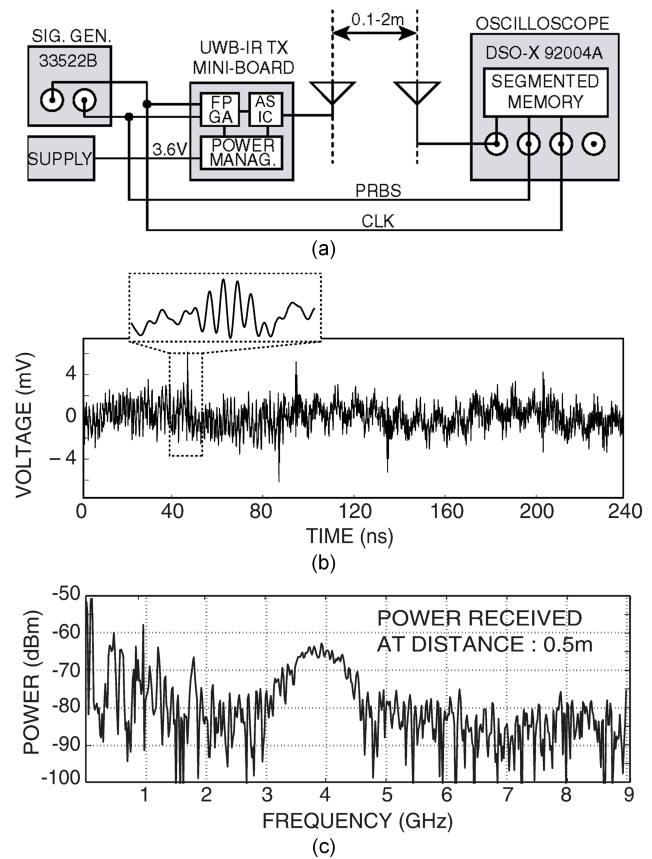}
 \caption{{\color{black}(a) Experimental setup for characterizing the developed UWB-IR TX in the {\color{black}wired-power wireless-data test} mode. (b) An UWB signal and (c) the corresponding power spectrum, {\color{black}both measured at} the RX antenna at a distance of 0.5 m.}}
 \label{fig_exp2}
 \end{figure}

Next, we tested the design in the wired-power wireless-data mode. {\color{black}This test mode allows us to characterize the performance of wireless transmission over the air without uncertainties in the power supply (since the device is powered through a wire). The data transmission was tested at different distances between the TX and RX modules.} The experimental setup of this test mode is illustrated in Fig.~\ref{fig_exp2} (a). The UWB receiver was the oscilloscope with an RX antenna. 
Similar to the previous setup (as shown in {\color{black}Fig.~\ref{fig_exp1} (a)}), the board was powered by an external supply and the biasing levels were generated by the on-chip DACs. The PRBS test data was fed to the TX board by the signal generator and was sent to the oscilloscope simultaneously for post-processing and BER computation. In the experiments with long distances (e.g., 1m and above), {\color{black}a synchronization signal was routed between the external power transmitter and the oscilloscope for triggering the segmented signal storage function of the oscilloscope. This is because the received pulses were too small to be detected by the notch trigger of the oscilloscope.}
Figs.~\ref{fig_exp2} (b) and (c) show the experimentally measured UWB signal and the corresponding power spectrum by the RX antenna at a distance of 0.5 m{\color{black}, respectively}.

\begin{figure}[!t]
 \centering
 \includegraphics[width=1\columnwidth]{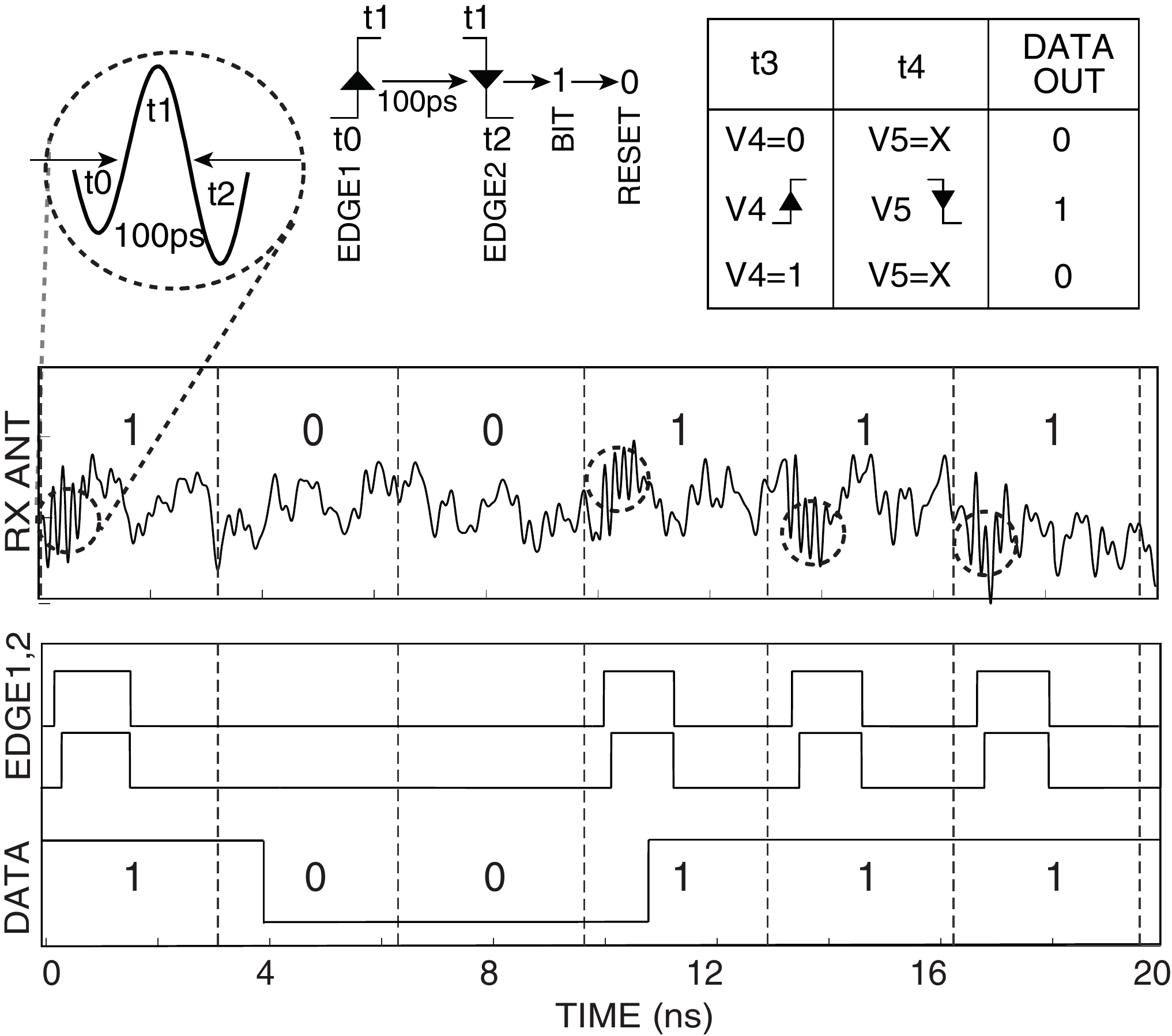}
 \caption{Measured output of the receiving antenna {\color{black}in the wireless-data test modes, as well as the} detected rising and falling edges and the data stream recovered by the detection algorithm.}
 \label{figrx_exp}
 \end{figure}

The receiver recovered the data by interpreting the measurements based on the scheme illustrated in Fig.~\ref{figrx_exp}. The oscilloscope was set to be triggered by any notch in the received signal that was narrower than 500ps. Once a notch was detected, a 10ns segment of the recorded RX signal containing the notch was stored in the memory. In an actual receiver circuit, the notch trigger can be replaced by an ultra-fast logic gate circuit that implements the 2-step edge detection scheme. 

The stored RX segments were processed offline to determine the BER. A correlated double sampling scheme was performed. By taking three consecutive samples from the RX signal in 100ps intervals, the algorithm detected whether a transmitted UWB pulse existed within each stored segment of the scope.
The ``EDGE1" and ``$\overline{\text{EDGE2}}$" signals were the outputs of two slope detection blocks which evaluated the rise in amplitude from $t0$ to $t1$, and from $t1$ to $t2$, respectively. A UWB pulse was flagged to be present within the segment when the output of both slope detectors ``EDGE1" and ``$\overline{\text{EDGE2}}$" were high.
A bit ``1" was assigned to each recorded RX segment when the algorithm detected a UWB pulse during that segment. Each segment also had a time stamp recorded using a separate channel. A bit ``0" was assumed where no pulse was detected by the algorithm. By comparing the bit ``1" segments with the original transmitted PRBS sequence, the BER was calculated. 




\subsection{\color{black}VCO-based All-Wireless Test Mode}

\begin{figure}[!ht]
 \centering
 \includegraphics[width=1\columnwidth]{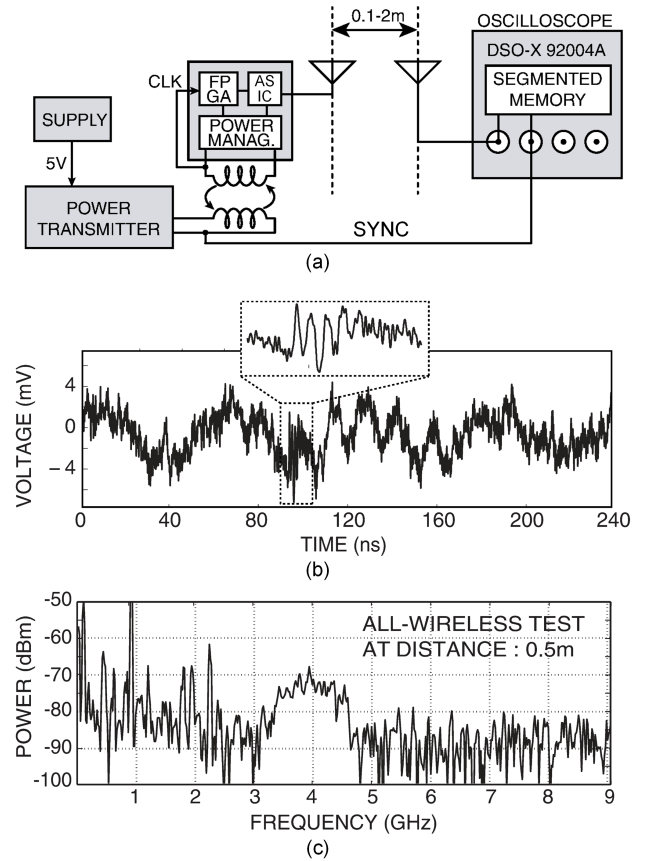}
 \caption{{\color{black} (a) Experimental setup for testing the UWB-IR in the {\color{black}VCO-based} all-wireless mode (inductively powered). (b) Experimentally measured UWB signal, and (c) the corresponding power spectrum {\color{black}measured at} the RX antenna in the {\color{black}VCO-based} all-wireless {\color{black}test} mode.}}
 \label{fig_exp3}
 \end{figure}

Lastly, we characterized the system in the VCO-based all-wireless mode. {\color{black}This test mode allows us to test the wireless data transmission while receiving power simultaneously. We benchmarked the final BER and power efficiency in this mode.} Fig.~\ref{fig_exp3} (a) illustrates the experimental setup for this all-wireless mode.
In this mode, the power management circuits receive power directly from the {\color{black}power receiver} (as shown in Fig.~\ref{figsimul}) {\color{black}and power the whole TX circuit}. The buck converter regulates its input (between 10V-40V) to a constant output of 3.6V, which powers two on-board regulators, one for the chip and the FPGA IO banks, the other for the FPGA core. 
A high-frequency choke isolates the chip's supply from the FPGA's supply to minimize the noise coupling.
Figs.~\ref{fig_exp3} (b) and (c) show the measurement results of the chip in the all-wireless mode at a distance of 0.5 m.

The TX used a 2cm$\times$2cm planar rectangular coil with the same footprint as the receiver. The coil was developed on a 2-layer flexible PCB substrate with turns on the top and bottom layers in series to increase the quality factor.
The inductive coil (positive) terminal was routed to the on-board FPGA, which generated the reference clock from the coil. To avoid shunting the coil current through the ESD protection diodes of the FPGA IO banks, a $30K\Omega$ resistor was placed between the coil and the FPGA (not shown in {\color{black}Fig.~\ref{fig_exp3} (a)} for simplicity).

\begin{figure}[!ht]
\centering
\includegraphics[width=2.8in]{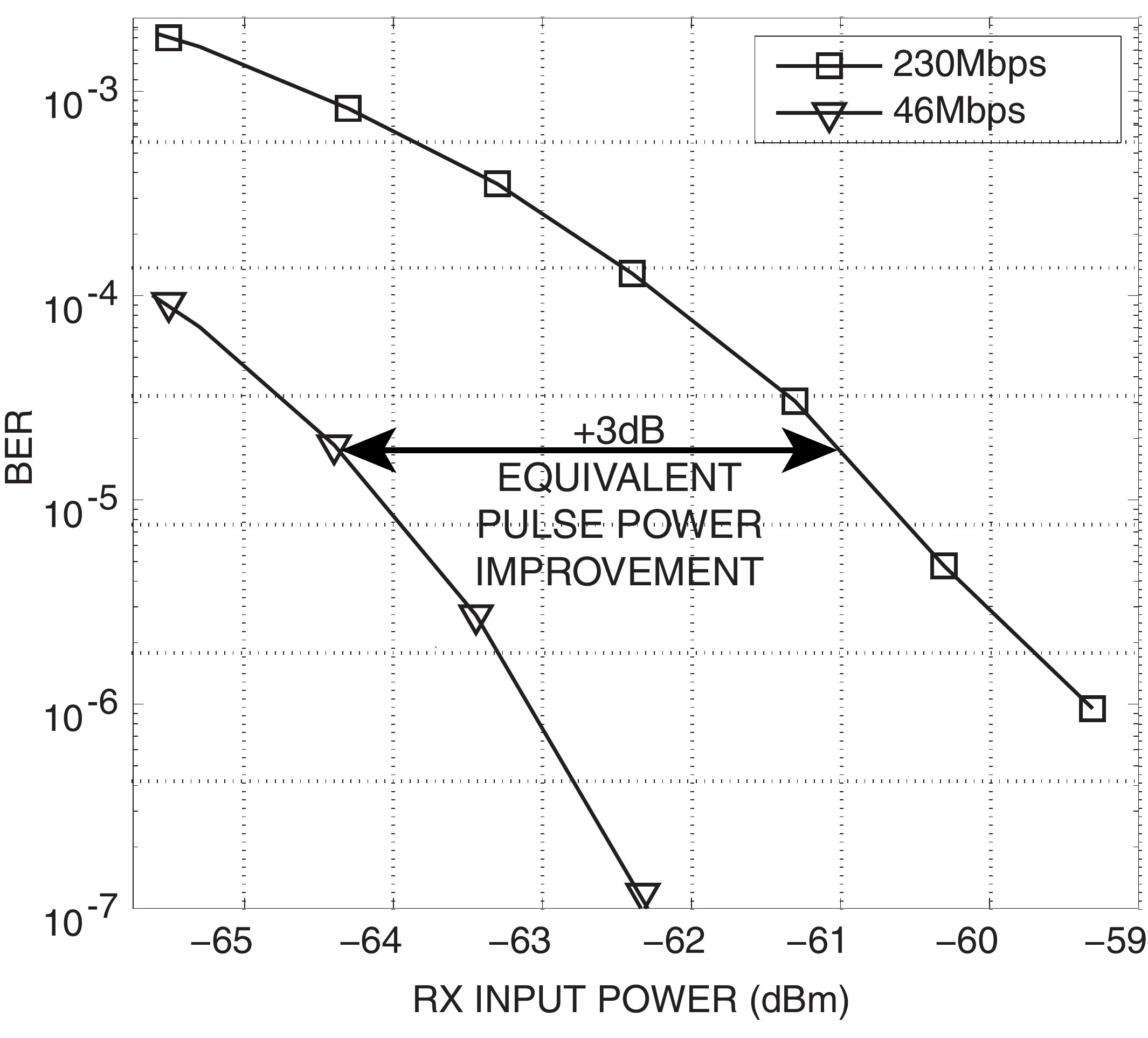}
\caption{Experimentally measured bit-error-rate (BER) {\color{black}in the VCO-based all-wireless mode}.}
\label{figBER}
\end{figure}

{\color{black}The power efficiency was measured in the VCO-based all-wireless test mode. When the output power is programmed to -1dBm and the chip is transmitting at 230Mbps, the TX power efficiency was measured to be 21.35\%.}
Fig.~\ref{figBER} shows the measured BER {\color{black}in the all-wireless test mode at a distance of 1m. During measurement, we swept the TX output power to collect the data points for different Rx input power levels. The RX input power was measured using the spectrum analysis function of the oscilloscope.}

{\color{black}We first tested device with a high data rate of 230Mbps. At a distance of 1m and a TX output power of -1dBm, the RX input power is about -59dBm, which includes the pass loss over the air and the antenna gains. The integrated noise of the RX over a bandwidth of 1.5GHz is about -70dBm, measured using a resolution bandwidth (RBW) of 100kHz. This yields a SNR of about 11dB, which is needed to achieve a $10^{-6}$ BER for OOK modulation \cite{Nikookar2009}. To improve the BER performance, we also tested a pulse averaging scheme. The averaging scheme uses multiple transmitted pulses to represent one bit, and the averaged power of the received consecutive pulses is used to determine the bit value \cite{singh2021}. We applied a 5-pulse averaging scheme to the 230Mbps transmission, resulting a reduced data rate of 46Mbps.}
{\color{black}The experimental results in Fig. \ref{figBER} show that the applied 5-pulse averaging scheme achieved a 3dB improvement for the BER. However, it should be noted that the energy per symbol for using 5 pulses averaging increases by 5x ($\sim$14dB). In conclusion, the pulse averaging scheme is not as efficient as directly increasing the power per pulse, which can be achieved by adjusting the clipping threshold voltages (i.e., $V_{MID}$ and $V_{MAX}$) in this work. However, increasing TX power is essentially limited by the total power budget of the device. For energy-constrained applications such as small medical devices, increasing TX power may not be feasible. In these cases, pulse averaging may be used as an alternative to improve the BER at the cost of additional power dissipation. } 


{\color{black}The inductive powering and data receiving were also characterized. The operating frequency of the inductive link is 1.5MHz. The frequency is chosen as it provides much stronger magnetic field compared to that at higher frequencies. Also, since the frequency is very low, the power transfer link causes negligible inference in the TX data transmission. The max power transfer efficiency was 28\% with a 4mA load current, and 40\% with a 10mA load current.}


\section{{\color{black}Discussion: Potential for Extension to VCO-free All-Wireless Mode}}

The proposed UWB-IR TX architecture has an additional advantage of being easily {\color{black}integratable} with resonant inductive power harvesting front-ends. The main design feature that eases the integration is {\color{black}the presence of} a resonant LC tank in both circuits. If the UWB-IR and the inductive power harvester are designed such that they share the same operating frequency, energy can be easily transferred between the two LC tanks simply by placing them near each other \cite{MIT}.

\begin{figure}[!ht]
\centering
\includegraphics[width=0.9\columnwidth]{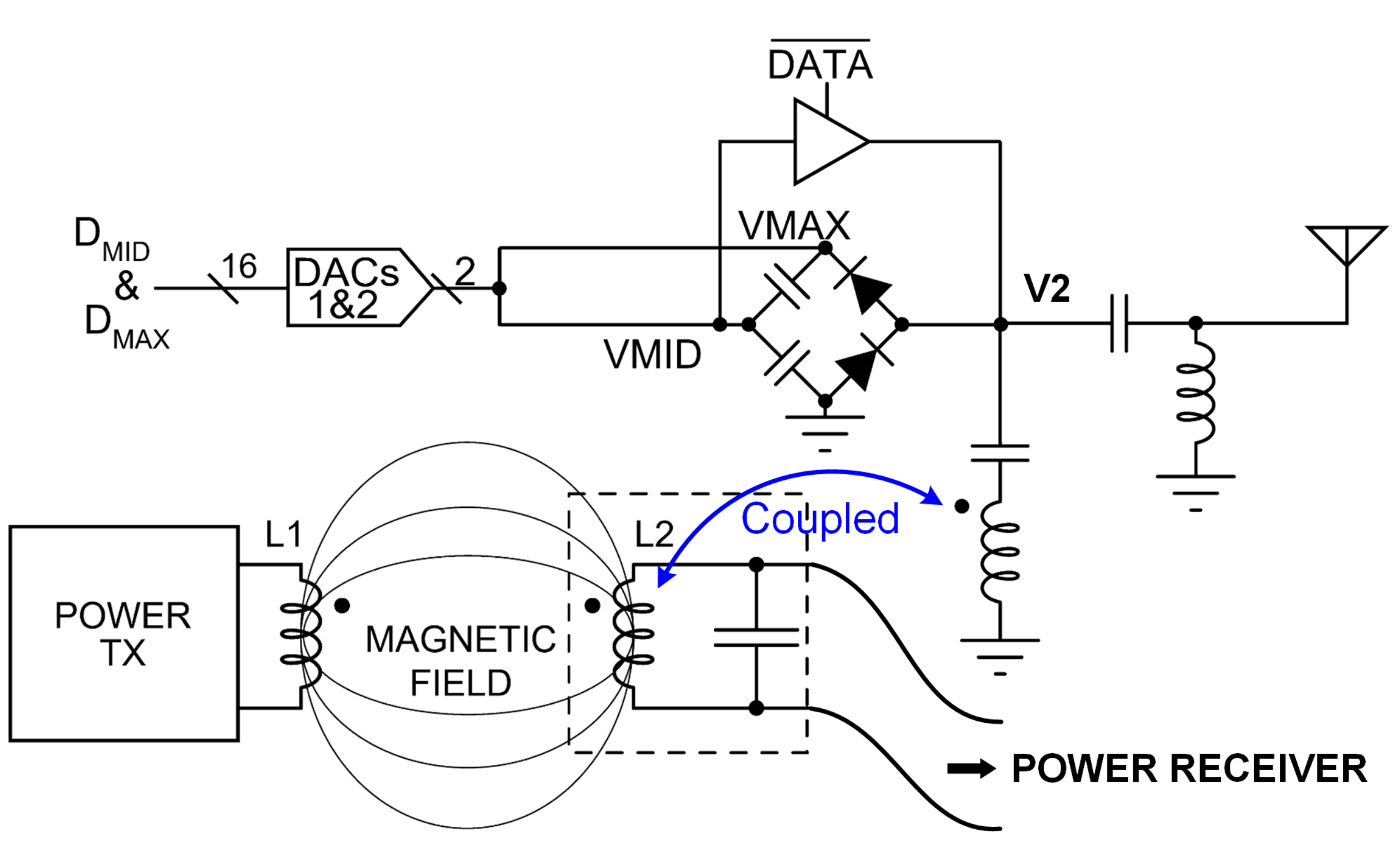}
\caption{{\color{black}A simplified diagram of the UWB-IR TX in the VCO-free all-wireless mode.}}
\label{fig16}
\end{figure}

Fig.~\ref{fig16} shows a block diagram of the UWB-IR TX when the LC tank in the TX receives power directly from the LC tank of the power receiver (Fig.~\ref{figsimul}). As shown in Fig.~\ref{fig16}, the frequency synthesizer, pre-amplifier, and the PA all can be removed if the LC tank receives power directly from the power receiver coil via magnetic coupling. This eliminates the dominant power consumption required to generate the high-precision tone to drive the LC tank, which would significantly further improve the power efficiency for low-power applications, such as energy-efficient sensors.

\begin{table*}[!ht]
\centering
\caption{\color{black}Comparative Analysis of Energy-efficient TX Designs}\label{table_comparison}
\renewcommand{\arraystretch}{1.3} 
\begin{tabu}{|[1pt]c|[1pt]cccc|[1pt]cccc|[1pt]}
\Xhline{1pt}
& \multicolumn{1}{c|}{\begin{tabular}[c]{@{}c@{}}JSSC '16\\ \cite{kim2016}\end{tabular}} & \multicolumn{1}{c|}{\begin{tabular}[c]{@{}c@{}}JSSC '17\\ \cite{streel2017}\end{tabular}} & \multicolumn{1}{c|}{\begin{tabular}[c]{@{}c@{}}JSSC '19\\ \cite{kopta2019}\end{tabular}} & \begin{tabular}[c]{@{}c@{}}JSSC '21\\ \cite{singh2021}\end{tabular} & \multicolumn{1}{c|}{\begin{tabular}[c]{@{}c@{}}JSSC '17\\ \cite{kassiri2017}\end{tabular}} & \multicolumn{1}{c|}{\begin{tabular}[c]{@{}c@{}}JSSC '19\\ \cite{lee2019}\end{tabular}}   & \multicolumn{1}{c|}{\begin{tabular}[c]{@{}c@{}}ISSCC '22\\ \cite{song2022}\end{tabular}}   & \begin{tabular}[c]{@{}c@{}}This \\ Work\end{tabular}       \\ \Xhline{1pt}
Architecture                                           & \multicolumn{4}{c|[1pt]}{\textbf{\color{black}Carrier-based}}                                                                                                                                                        & \multicolumn{4}{c|[1pt]}{\textbf{\color{black}Pulse-radio}}                                                                                                                                                                                                                                                                 \\ \hline
Modulation                                             & \multicolumn{1}{c|}{BPSK}     & \multicolumn{1}{c|}{BPSK}     & \multicolumn{1}{c|}{FM}                                                   & BPSK                                                   & \multicolumn{1}{c|}{PPM}                                                    & \multicolumn{1}{c|}{MPPM}                                                     & \multicolumn{1}{c|}{Hybrid}                                                    & OOK                                                        \\ \hline
CMOS Process                                           & \multicolumn{1}{c|}{130nm}    & \multicolumn{1}{c|}{28nm}     & \multicolumn{1}{c|}{65nm}                                                 & 28nm                                                   & \multicolumn{1}{c|}{130nm}                                                  & \multicolumn{1}{c|}{65nm}                                                     & \multicolumn{1}{c|}{28nm}                                                      & 130nm                                                      \\ \hline
Area (mm2)                                             & \multicolumn{1}{c|}{4.6}      & \multicolumn{1}{c|}{0.93}     & \multicolumn{1}{c|}{1.1}                                                  & 0.154                                                  & \multicolumn{1}{c|}{0.5}                                                    & \multicolumn{1}{c|}{2.88}                                                     & \multicolumn{1}{c|}{0.16}                                                      & 1.19                                                       \\ \hline
Supply                                                 & \multicolumn{1}{c|}{1V}       & \multicolumn{1}{c|}{+/-1.8V}  & \multicolumn{1}{c|}{1V}                                                   & 0.9V                                                   & \multicolumn{1}{c|}{0.5V}                                                   & \multicolumn{1}{c|}{1.1V}                                                     & \multicolumn{1}{c|}{-}                                                         & 1.2/3.3V                                                   \\ \hline
Bandwidth (GHz)                                        & \multicolumn{1}{c|}{7}        & \multicolumn{1}{c|}{-}        & \multicolumn{1}{c|}{0.5}                                                  & 1                                                      & \multicolumn{1}{c|}{1.25}                                                   & \multicolumn{1}{c|}{2}                                                        & \multicolumn{1}{c|}{-}                                                         & 2                                                          \\ \hline
Power (mW)                                             & \multicolumn{1}{c|}{22.6}     & \multicolumn{1}{c|}{0.65}     & \multicolumn{1}{c|}{0.575}                                                & 4.9                                                    & \multicolumn{1}{c|}{0.47}                                                   & \multicolumn{1}{c|}{7}                                                        & \multicolumn{1}{c|}{9.69}                                                      & 3.7                                                        \\ \hline
Pout (dBm)                                             & \multicolumn{1}{c|}{-8.7}     & \multicolumn{1}{c|}{-20}      & \multicolumn{1}{c|}{-11.4}                                                & -2.5                                                   & \multicolumn{1}{c|}{-12.6}                                                  & \multicolumn{1}{c|}{-6.35}                                                    & \multicolumn{1}{c|}{0}                                                         & -1                                                         \\ \hline
Data-rate (Mbps)                                       & \multicolumn{1}{c|}{1000}     & \multicolumn{1}{c|}{27.24}    & \multicolumn{1}{c|}{0.1}                                                  & 6.81                                                   & \multicolumn{1}{c|}{20}                                                     & \multicolumn{1}{c|}{500}                                                      & \multicolumn{1}{c|}{1660}                                                      & 230/46                                                     \\ \hline
Energy/bit (pJ)                                        & \multicolumn{1}{c|}{102.2}    & \multicolumn{1}{c|}{14}       & \multicolumn{1}{c|}{-}                                                    & 1.12                                                   & \multicolumn{1}{c|}{2.76}                                                   & \multicolumn{1}{c|}{4.7}                                                      & \multicolumn{1}{c|}{5.8}                                                       & 21                                                         \\ \hline
Distance (mm)                                          & \multicolumn{1}{c|}{1000}     & \multicolumn{1}{c|}{50-300}   & \multicolumn{1}{c|}{-}                                                    & -                                                      & \multicolumn{1}{c|}{100}                                                    & \multicolumn{1}{c|}{5000}                                                     & \multicolumn{1}{c|}{\color{black}150*}                                                        & {2000}                                              \\ \hline
{\color{black}BER} & \multicolumn{1}{c|}{{\color{black}$10^{-3}$}}     & \multicolumn{1}{c|}{{\color{black}-}}   & \multicolumn{1}{c|}{{\color{black}$10^{-3}$}}                                                    & {\color{black}-}                                                      & \multicolumn{1}{c|}{{\color{black}-}}                                                    & \multicolumn{1}{c|}{{\color{black}$10^{-3}$}}                                                     & \multicolumn{1}{c|}{{\color{black}$10^{-4}$}}                                                        & {\color{black}$10^{-6}$\dag}                                        
     \\ \hline
Wireless Powering                                      & \multicolumn{1}{c|}{No}       & \multicolumn{1}{c|}{No}       & \multicolumn{1}{c|}{No}                                                   & No                                                     & \multicolumn{1}{c|}{Yes}                                                    & \multicolumn{1}{c|}{No}                                                       & \multicolumn{1}{c|}{No}                                                        & {Yes}                                               \\ \hline
TX Efficiency                                          & \multicolumn{1}{c|}{0.59\%}   & \multicolumn{1}{c|}{2.6\%}    & \multicolumn{1}{c|}{12.2\%}                                               & 4.3\%                                                  & \multicolumn{1}{c|}{11.7\%}                                                 & \multicolumn{1}{c|}{3.29\%}                                                   & \multicolumn{1}{c|}{10.32\%}                                                   & {21.35\%}                                           \\ \Xhline{1pt}
\end{tabu}\\ \vspace{1.5mm}
\hspace{1.5cm} \raggedright \footnotesize{\color{black}*This distance includes a 15 mm thick tissue and an implant antenna.\\}
\hspace{1.5cm} \raggedright \footnotesize{\dag This BER was measured with 230Mbps at 1m. It can also be achieved at 2m with a 46Mbps data rate.}

\end{table*}

One limitation of using direct-coupled LC tanks is that the power coil must resonate at the same frequency as the UWB-IR TX. The TX pulse generation frequency, however, may not always be the optimal frequency for the power transfer system. The operating frequency of inductive power transfer systems is often selected at the lower MHz frequency (usually 13.67MHz ISM band) \cite{liu2016}. Higher MHz operating frequencies are often avoided because the coils become increasingly radiative at these high frequencies, the quality factor is more limited, and the allowable magnetic field intensity is also more limited \cite{ieeec95}. Therefore, when directly coupled to an inductive power receiving inductor, the maximum practical data-rate of the UWB-IR TX is about 20Mbps-30Mbps. However, even this reduced data-rate is still sufficient for many sensory microsystems.

\begin{figure}[!ht]
\centering
\includegraphics[width=3in]{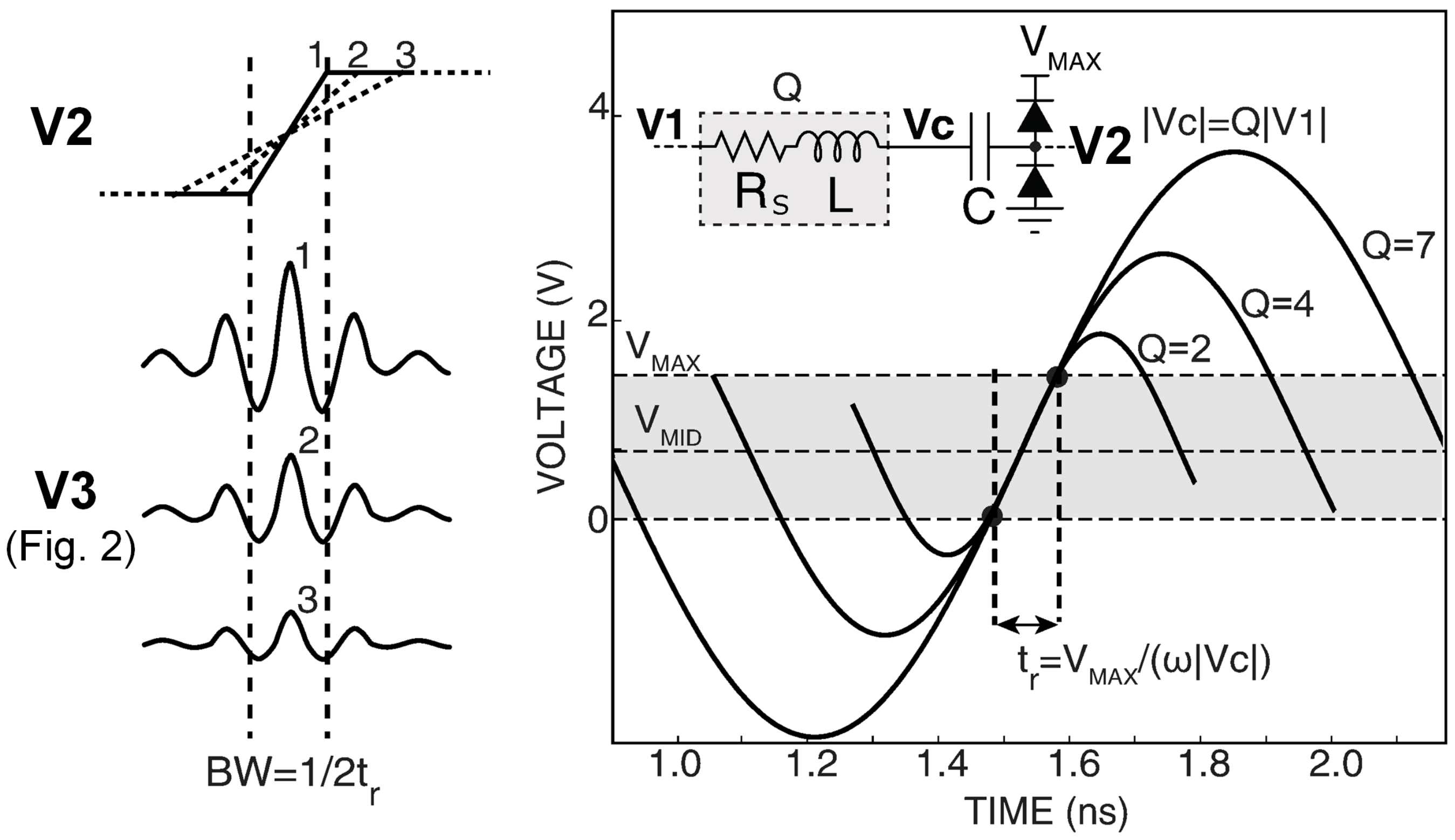}
\caption{Illustration of the increased requirement in quality factor when the LC tank is coupled to a low-frequency inductive resonator.}
\label{figswing}
\end{figure}

{\color{black}Another} challenge in sourcing the driving signal of the UWB-IR TX LC tank from the inductive power receiving coil is the lower rising and falling times of the clipped signal $V_2$ for the same center node voltage $V_c$ swing of the LC tank. This is again caused by the lower oscillation frequency of the tank, which is set by the inductive power transfer frequency. Fig.~\ref{figswing} shows the relationship between the rising and falling times of the clipped signal $V_2$, the operating frequency, and the center node voltage swing $V_c$. When the operating frequency is lowered, the quality factor of the LC tank must be increased proportionally. The increased quality factor and signal swing ensure that the rising time of the clipped signal $V_2$ remains the same, and so does the power and bandwidth of the generated UWB pulse, despite the lowered operating frequency. The UWB-IR inductor should have a similar design and dimensions as the power coil, since the inductive powering coil also needs to be designed for the highest possible quality factor given the geometry constraints.

\section{\color{black}Comparative Analysis}
\begin{figure}[!ht]
\centering
\includegraphics[width=0.95\columnwidth]{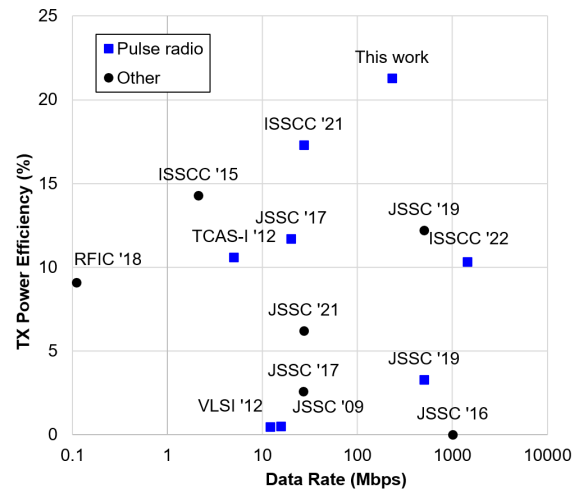}
\caption{Comparison with prior art {\color{black}in terms of the} TX power efficiency and the data rate. }
\label{figcomp_comp}
\end{figure}

Table \ref{table_comparison} compares key specifications of the presented work with the state-of-the-art TX designs, including both pulse-radio and carrier-based architectures. Fig.~\ref{figcomp_comp} plots the power efficiency as a function of the data rate for this and other existing designs.
As discussed in Section I, the output power efficiency of existing UWB-IR TX architectures is low, leading to a limited operating range. {\color{black}Fig. 17 compares the TX power efficiency and data rate of the state-of-the-art and the proposed design. It is challenging to achieve both high power efficiency and high data rate simultaneously, as high data rate often comes at the cost of circuit complexity and power overhead, such as a more complicated modulation scheme \cite{song2022}. The proposed work represents a good trade-off and design improvement in both dimensions.}
Owing to the proposed clipped-sinusoid pulse generation scheme, the presented work shows a TX power efficiency of 21.3\%, at a data-rate of 230Mb/s, which is the highest {\color{black}reported power efficiency, even when including both carrier-based narrow-band TXs and pulse radios, as shown in Table I, on the left and right panels, respectively}. Unlike several designs {\color{black}operating at} short transfer distances \cite{song2022,kassiri2017,streel2017}, {\color{black}this design's} performance was achieved at an over-the-air distance of up to 2m. This longer distance is essential for many biomedical sensors and body-area sensor networks applications \cite{liu2021}, including electronic neural interfaces in freely behaving animals \cite{Nima2016}. Finally, the energy consumption is calculated to be 21pJ/b, which is partially limited by the {\color{black}conservative} 130nm CMOS technology used for prototyping. {\color{black}Although the design in [18] was able to achieve a lower energy per bit performance using the same CMOS node, it was designed to operate at a distance of 100mm, which is 20x shorter than this work.} In a technology with smaller feature size, a lower supply voltage should yield further reduced energy consumption. {\color{black}In addition, techniques such as phase scrambling can be further used to reduce the spurious tones in the TX output spectrum. The corresponding small power consumption overhead due to such an implementation may result in a SNR improvement that directly benefits the BER.}

\section{Conclusion}
\label{conclusion}

In this paper, {\color{black}a power-efficient clipped-sinusoid UWB-IR TX design was presented. Simultaneous powering and data transmission is supported. A 130nm prototype was fabricated and tested. A state-of-the-art TX power efficiency of 21.3\% at a data-rate of 230Mb/s has been achieved. }
A BER of less was $10^{-6}$ was measured at 46Mbps over a distance of 2m, and 230Mbps over a distance of 1m. 
The energy consumption is 21pJ/b. 
{\color{black}Additionally, the design can be configured in a VCO-free all-wireless mode for even lower power low-data-rate applications.}
The proposed {\color{black}clipped-sinusoid UWB-IR TX design} can be {\color{black}deployed} in a broad range of applications, especially in power-constrained biomedical wearable and implantable sensory microsystems with high data rate requirements or {\color{black}when} wireless powering and data receiving are desirable.

\ifCLASSOPTIONcaptionsoff
  \newpage
\fi

\bibliographystyle{IEEEtran}
\bibliography{ref}

\ifCLASSOPTIONcaptionsoff
  \newpage
\fi

\begin{IEEEbiography}[{\includegraphics[width=1in,height=1.25in,clip,keepaspectratio]{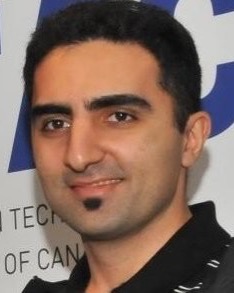}}]{Nima Soltani} (Member, IEEE) received the B.Eng. and M.A.Sc. degrees in electrical engineering from Ryerson University, Toronto, ON, Canada, in 2007 and 2010, respectively, and the Ph.D. degree from the University of Toronto, Toronto, in 2016. His M.A.Sc. thesis focused on ultra low-power RF circuits for passive microsystems and RF power transmission. His doctoral dissertation focuses on inductively-powered implantable brain chemistry monitoring systems. From 2010 to 2011, he was with Solace Power Inc., Mount Pearl, NL, Canada, on the development of the first electrical induction system for medium-range wireless power transfer. He is the Co-Founder of BrainCom Inc., Toronto, a Toronto-based company specializing in the development of wearable and implantable platforms for high-resolution brain signal processing and acquisition. He is currently a Senior Analog Designer with Intel Corporation, Canada.
\end{IEEEbiography}

\begin{IEEEbiography}[{\includegraphics[width=1in,height=1.25in,clip,keepaspectratio]{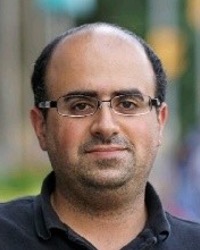}}]{Hamed Mazhab Jafari} received the B.Eng. and M.A.Sc. (focused on low-power UWB CMOS front ends and UWB antennas) degrees in electrical engineering from McMaster University, Hamilton, ON, Canada, in 2004 and 2006, respectively, and the Ph.D. degree in electrical and computer engineering from the University of Toronto, Toronto, ON, Canada, in 2013. He has held Internship positions with Kapik Integration, Toronto, ON, Canada, where he worked on low-power mixed-signal circuits. Between 2011 and 2015, he was with Snowbush IP, Toronto, ON, Canada, where he focuses on the research and development of next-generation high-speed wireline communication systems. He is currently an Analog and Mixed-signal Designer with Intel Corporation, Canada.
\end{IEEEbiography}

\begin{IEEEbiography}[{\includegraphics[width=1in,height=1.25in,clip,keepaspectratio]{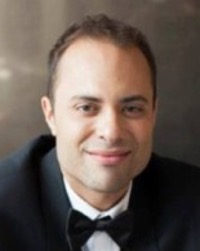}}]{Karim Abdelhalim} received the B.Eng. and M.A.Sc. degrees in electrical engineering from Carleton University, Ottawa, ON, Canada, in 2005 and 2007, respectively, and the Ph.D. degree in 2013 in electrical and computer engineering from the University of Toronto, Toronto, ON, Canada, where he focused on wireless neural recording and stimulation SoCs and their application in monitoring and treatment of intractable epilepsy. From 2011 to 2015, he was a Senior Staff Scientist at Broadcom Corporation, Irvine, CA, USA, where he was involved with the design of mixed-signal ICs for 10/100/1G-BASE-T Ethernet applications and 100BASE-T1 and 1000BASE-T1 automotive Ethernet applications. From July 2010 to October 2010, he also worked as a mixed-signal design engineering intern at Broadcom, Irvine, CA, USA. From 2015 to 2017, he was a Principal Engineer at Inphi Corporation where he was involved with the design of high-speed optical transceivers operating at 28GS/s and 56GS/s in 16nm CMOS and from 2017 to 2022 he was a Director of Engineering at Sitrus Technology Corporation and was involved in the design and production of 28Gb/s and 56Gb/s optical transceivers. He returned to Broadcom’s Physical Layer Products (PLP) as a mixed-signal IC designer focusing on automotive Ethernet, high-speed SERDES and optical applications.
\end{IEEEbiography}

\begin{IEEEbiography}[{\includegraphics[width=1in,height=1.25in,clip,keepaspectratio]{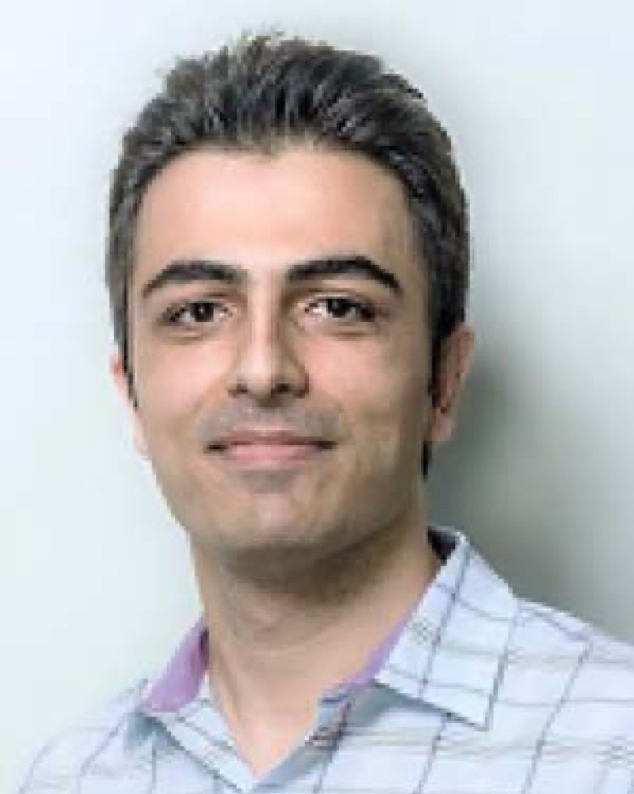}}]{Hossein Kassiri} received the B.Sc. degree in electrical and computer engineering from the University of Tehran, Tehran, Iran, in 2008, M.A.Sc. degree in electrical and computer engineering from McMaster University, Hamilton, ON, Canada, in 2010, and Ph.D. degree in electrical and computer engineering from the University of Toronto, Toronto, ON, Canada, in 2015. He is currently an Associate Professor with the Department of Electrical Engineering and Computer Science, York University, Toronto, ON, where he is also the Director of the Integrated Circuits and Systems Laboratory and the Center for Microelectronics Prototyping and Test. In September 2015, he Co-Founded BrainCom Inc., which specialized in implantable brain-computer interfaces. His research interests include the area of design and development of wireless and battery-less multi-modal neural interfacing systems and their application in the monitoring and treatment of neurological disorders.

Dr. Kassiri is the recipient of the IEEE BioCAS 2021 Best paper award, IEEE ISSCC 2017 Jack Kilby Award for Outstanding Student Paper, IEEE ISCAS Best Paper Award - BioCAS track (2016), Ontario Brain Institute Entrepreneurship Award in 2015, Heffernan Commercialization Award in 2014, and the CMC Brian L. Barge Award for Excellence in Microsystems Integration in 2012.
\end{IEEEbiography}

\begin{IEEEbiography}[{\includegraphics[width=1in,height=1.25in,clip,keepaspectratio]{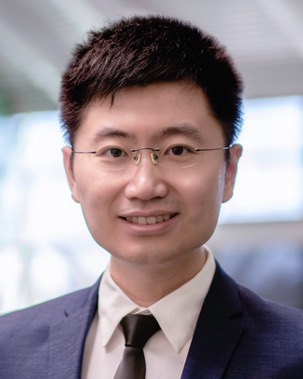}}]{Xilin Liu} (Senior Member, IEEE) obtained his Ph.D. degree from the University of Pennsylvania, Philadelphia, PA, USA, in 2017. 

He is currently an Assistant Professor of Electrical and Computer Engineering at the University of Toronto and an affiliated scientist at the University Health Network (UHN), Toronto, ON, Canada. His research interests include analog and mixed-signal IC design for emerging applications in healthcare and communication. Before joining the University of Toronto in 2021, he held industrial positions at Qualcomm Inc., where he conducted research and development of high-performance mixed-signal circuits for cellular communication. He led and contributed to the IPs that have been integrated into products in high-volume production. He was a visiting scholar at Princeton University in 2014.

Dr. Liu received the Best Student Paper Award and the Best Track Award at the 2017 ISCAS, the Best Paper Award (1st place) at the 2015 BioCAS, the Best Track Award at the 2014 ISCAS, and the student research preview (SRP) award at the 2014 ISSCC. He also received the SSCS Predoctoral Achievement Award at the 2016 ISSCC.
\end{IEEEbiography}

\begin{IEEEbiography}[{\includegraphics[width=1in,height=1.25in,clip,keepaspectratio]{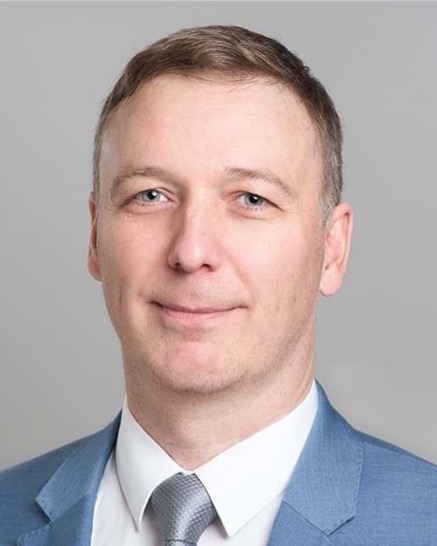}}]{Roman Genov} (Senior Member, IEEE) received the B.S. degree in Electrical Engineering from Rochester Institute of Technology, NY in 1996 and the M.S.E. and Ph.D. degrees in Electrical and Computer Engineering from Johns Hopkins University, Baltimore, MD in 1998 and 2003 respectively.

He is currently a Professor in the Department of Electrical and Computer Engineering at the University of Toronto, Canada, where he is a member of Electronics Group and Biomedical Engineering Group and the Director of Intelligent Sensory Microsystems Laboratory. Dr. Genov’s research interests are primarily in analog and digital integrated circuits and systems for energy-constrained biomedical and consumer sensory applications, such as implantable neural interfaces and computational image sensors.

Dr. Genov is a co-recipient of Jack Kilby Award for Outstanding Student Paper at IEEE International Solid-State Circuits Conference, Best Paper Award of IEEE TRANSACTIONS ON BIOMEDICAL CIRCUITS AND SYSTEMS, Best Paper Award of IEEE Biomedical Circuits and Systems Conference, Best Student Paper Award of IEEE International Symposium on Circuits and Systems, Best Paper Award of IEEE Circuits and Systems Society Sensory Systems Technical Committee, Best Paper Award of IEEE Circuits and Systems Society Biomedical Circuits and Systems Technical Committee, GlobalFoundries Micro-Nanosystems Design Award, Award for Excellence in Microsystems Design Methodology, Brian L. Barge Award for Excellence in Microsystems Integration, MEMSCAP Microsystems Design Award, DALSA Corporation Award for Excellence in Microsystems Innovation, and Canadian Institutes of Health Research Next Generation Award. He was a Technical Program Co-chair at IEEE Biomedical Circuits and Systems Conference, a member of IEEE International Solid-State Circuits Conference International Program Committee, and a member of IEEE European Solid-State Circuits Conference Technical Program Committee. He was also an Associate Editor of IEEE TRANSACTIONS ON CIRCUITS AND SYSTEMS-II: EXPRESS BRIEFS and IEEE SIGNAL PROCESSING LETTERS, as well as a Guest Editor for IEEE JOURNAL OF SOLID-STATE CIRCUITS. Currently he is an Associate Editor of IEEE TRANSACTIONS ON BIOMEDICAL CIRCUITS AND SYSTEMS.

\end{IEEEbiography}

\end{document}